%% file: main.tex
\documentclass[a4paper,11pt]{article}
\pdfoutput=1 

\usepackage{jheppub} 
\usepackage[utf8]{inputenc}
\usepackage{siunitx}
\usepackage{xcolor,colortbl,listings}

\usepackage{chngcntr}
\counterwithin{footnote}{section} 

\usepackage{amssymb}
\usepackage{array}
\usepackage{lscape}
\usepackage[version=4]{mhchem}
\usepackage[noabbrev,capitalise]{cleveref}
\usepackage{eurosym}
\usepackage{xspace}
\usepackage[section]{placeins}
\usepackage{longtable}
\usepackage{tabularx}
\usepackage{afterpage}
\usepackage{float}
\usepackage{lscape}
\usepackage{subfigure}
\usepackage{wrapfig}
\usepackage{lastpage}
\usepackage{fancyhdr}
\usepackage{makecell}
\usepackage{placeins}

\usepackage{bbding,upgreek,pifont,commath,wasysym,fourier-orns}
\usepackage{enumitem}
\usepackage[super]{nth}
\usepackage[export]{adjustbox}
\usepackage[b]{esvect}
\usepackage{environ,indentfirst}

\usepackage{comment}

\usepackage{tikz}
\usetikzlibrary{patterns}
\usetikzlibrary{plotmarks}
\usetikzlibrary{external}

\usetikzlibrary{positioning, arrows, shapes, calc}
\tikzset{->, >=stealth',
sq/.style={draw, circle},
iq/.style={draw, ellipse}
}

\usepackage{amssymb,amsmath,mathtools,cases,enumitem}

\usepackage{booktabs,multirow,ragged2e}

\newcolumntype{L}[1]{>{\raggedright\let\newline\\\arraybackslash\hspace{0pt}}m{#1}}
\newcolumntype{C}[1]{>{\centering\let\newline\\\arraybackslash\hspace{0pt}}m{#1}}
\newcolumntype{R}[1]{>{\raggedleft\let\newline\\\arraybackslash\hspace{0pt}}m{#1}}
\newcolumntype{N}{@{}m{0pt}@{}}

\newcommand{\lsim}{\mathrel{\mathop{\kern 0pt \rlap
  {\raise.2ex\hbox{$<$}}}
  \lower.9ex\hbox{\kern-.190em $\sim$}}}
\newcommand{\gsim}{\mathrel{\mathop{\kern 0pt \rlap
  {\raise.2ex\hbox{$>$}}}
  \lower.9ex\hbox{\kern-.190em $\sim$}}}

\newcommand{\alt}{\mathrel{\mathop{\kern 0pt \rlap
  {\raise.2ex\hbox{$<$}}}
  \lower.9ex\hbox{\kern-.190em $\sim$}}}
\newcommand{\agt}{\mathrel{\mathop{\kern 0pt \rlap
  {\raise.2ex\hbox{$>$}}}
  \lower.9ex\hbox{\kern-.190em $\sim$}}}

\newcommand{\gagamma}{g_{a\gamma}}

\newcommand{\gae}{g_{ae}}
\newcommand{\dd}{\mathrm{d}}

\newcommand{\exclude}[1]{}

\title{An accurate solar axions ray-tracing response of BabyIAXO}

\collaboration{
    \includegraphics[height=17mm]{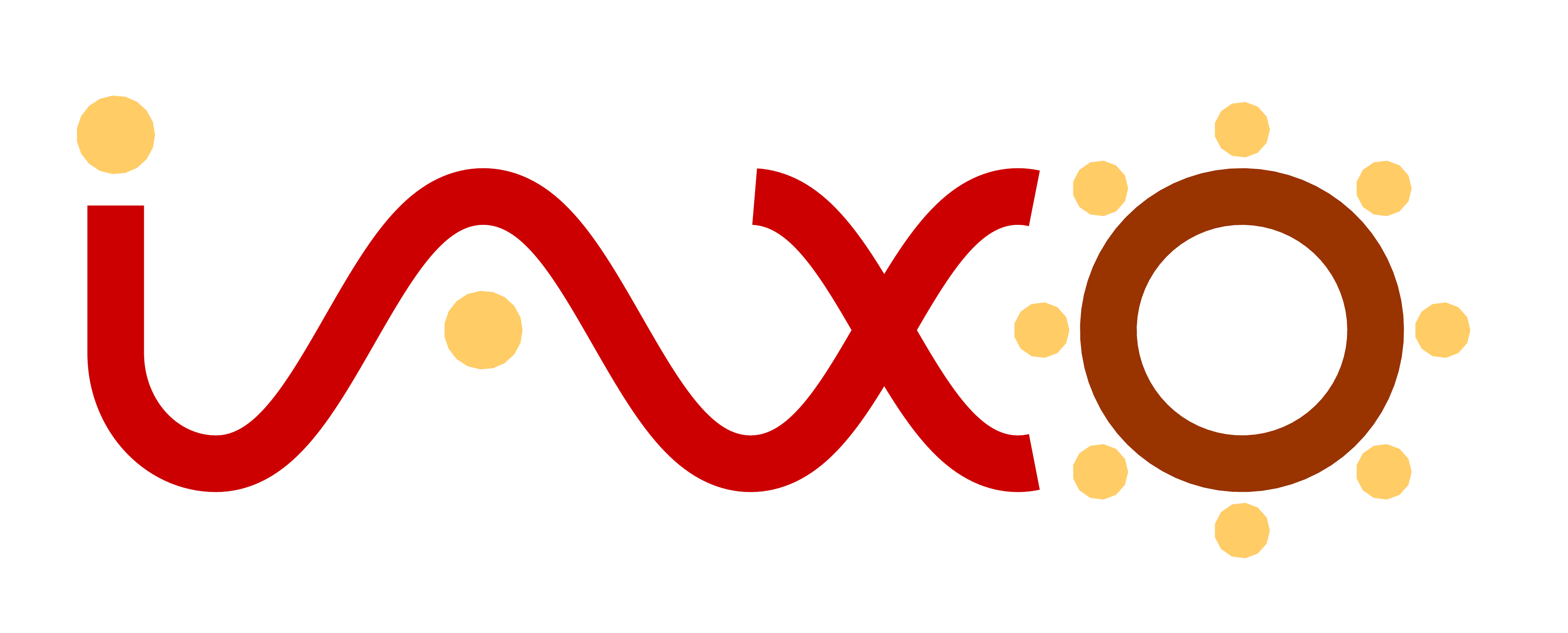}\\ [6pt] IAXO collaboration}

\input{IAXO_Authors_JHEP}

\emailAdd{javier.galan@unizar.es}

\abstract{

BabyIAXO is the intermediate stage of the International Axion Observatory (IAXO) to be hosted at DESY. Its primary goal is the detection of solar axions following the axion helioscope technique. Axions are converted into photons in a large magnet that is pointing to the sun. The resulting X-rays are focused by appropriate X-ray optics and detected by sensitive low-background detectors placed at the focal spot. 
The aim of this article is to provide an accurate quantitative description of the different components (such as the magnet, optics, and X-ray detectors) involved in the detection of axions. 
Our efforts have focused on developing robust and integrated software tools to model these helioscope components, enabling future assessments of modifications or upgrades to any part of the IAXO axion helioscope and evaluating the potential impact on the experiment's sensitivity. In this manuscript, we demonstrate the application of these tools by presenting a precise signal calculation and response analysis of BabyIAXO's sensitivity to the axion-photon coupling. Though focusing on the Primakoff solar flux component, our virtual helioscope model can be used to test different production mechanisms, allowing for direct comparisons within a unified framework.
}

\begin{document}

\maketitle

\section{Introduction}
\label{sec:intro}
\input{sections/01_introduction.tex}


\input{sections/02_components.tex}



\input{sections/03_sensitivity.tex}







\input{sections/05_conclusions.tex}

\section*{Acknowledgments} 

We acknowledge support from the European Research Council (ERC) under the European Union’s Horizon 2020 research and innovation programme, grant agreements ERC-2017-AdG-788781 (IAXO+) and ERC-2018-StG-802836 (AxScale). 

We acknowledge support from the Deutsches Elektronen-Synchrotron (DESY), Hamburg, Germany, a member of the Helmholtz Association (HGF). This research was realized in part on the National Analysis Facility (NAF) computational resources operated at DESY.

We acknowledge funding from the Agencia Estatal de Investigación (AEI) under grants PID2019-108122GB and PID2022-137268NB funded by MCIN/AEI/10.13039/501100011033 and by ''ERDF – A way of making Europe''. Additional support is provided through the ''European Union NextGenerationEU/PRTR'' (Planes Complementarios, Programa de Astrofísica y Física de Altas Energías) and the DGA-FSE grants 2023-E21-23R, 2023-E27-23R and 2023-E16-23R. We acknowledge support from INAF, the Italian National Institute for Astrophysics, through the large grant ``BRAVO SUN''. We gratefully acknowledge support from the Agence Nationale de la Recherche (France) under grant ANR-19-CE31-0024. 

We acknowledge support from the US Department of Energy (DOE) under Contract No. DE-AC52-07NA27344, performed at the Lawrence Livermore National Laboratory. We also acknowledge support from the National Science Foundation (NSF), Award \#2309980.  

We are grateful for the support from the South African Research Chairs Initiative of the Department of Science and Technology and the National Research Foundation of South Africa. We also acknowledge support from the ICTP Associates Programme and the Simons Foundation through grant number 284558FY19.

Funding has also been received from CA21106 Cosmic WISPers. SH acknowledges support from the European Union’s Horizon Europe research and innovation programme under the Marie Skłodowska-Curie grant agreement No. 101065579


\clearpage
\appendix

\input{sections/appendixA}

\input{sections/appendixB}


\input{sections/appendixD}

\input{sections/appendixC}


\bibliographystyle{JHEP}
\bibliography{bib/IAXObib,bib/detectors,bib/new,bib/theory}


\end{document}

%% file: IAXO_Authors_JHEP.tex
\newcommand{\UNIZARname}{Centro de Astropart\'iculas y F\'isica de Altas Energ\'ias (CAPA), Universidad de Zaragoza, 50009 Zaragoza, Spain}
\newcommand{\CEFCAname}{Centro de Estudios de F\'{\i}sica del Cosmos de Aragon, Plaza San Juan, Teruel, Spain}
\newcommand{\ICCUBname}{Departament de F\'isica Qu\`antica i Astrofis\'ica \& Institut de Ci\`encies del Cosmos, Universitat de Barcelona, Barcelona, Spain}
\newcommand{\ICREAname}{Instituci\'o Catalana de Recerca i Estudis Avançats, ICREA, Barcelona, Spain}
\newcommand{\Mainzname}{Institute of Physics, Johannes Gutenberg University, Mainz, Germany}
\newcommand{\PNPIname}{Petersburg Nuclear Physics Institute - NRC Kurchatov Institute, Gatchina, Russian Federation}
\newcommand{\RBIname}{Rudjer Bo\v{s}kovi\'{c} Institute, Bijeni\v{c}ka cesta 54, 10000 Zagreb, Croatia}
\newcommand{\UBonnname}{Physikalisches Institut der Universit\"at Bonn, Nussallee 12, 53115 Bonn, Germany}
\newcommand{\UCTname}{High Energy Physics, Cosmology \& Astrophysics Theory (HEPCAT) group, University of Cape Town, Private Bag, 7700 Rondebosch, South Africa}
\newcommand{\AIMSname}{African Institute for Mathematical Sciences, 6 Melrose Road, Muizenberg, 7945, South Africa}
\newcommand{\DESYname}{Deutsches Elektronen-Synchrotron DESY, Notkestr.\,85, 22607 Hamburg, Germany}
\newcommand{\CERNname}{CERN - European Organization for Nuclear Research, Geneva, Switzerland}
\newcommand{\UHEIname}{Heidelberg University, Kirchhoff Institute for Physics}
\newcommand{\LLNLname}{Lawrence Livermore National Laboratory, Livermore, CA, U.S.A.}
\newcommand{\Columbianame}{Columbia University, Columbia Astrophysics Laboratory, New York, NY U.S.A.}

\newcommand{\INAFTname}{INAF, Osservatorio Astronomico d'Abruzzo, Via Mentore Maggini, Teramo, Italy}

\newcommand{\INFNRname}{INFN, Istituto Nazionale di Fisica Nucleare, Sezione di Roma, Italy}

\newcommand{\VATICANname}{Specola Vaticana (Vatican Observatory), V-00120, Vatican City State}

\newcommand{\INAFBname}{INAF, Osservatorio di Astrofisica e Scienza dello spazio, via Gobetti 101, I-40129 Bologna, Italy}

\newcommand{\INFNBname}{INFN, Istituto Nazionale di Fisica Nucleare, Sezione di Bologna, via Irnerio 46, 40126 Bologna, Italy}

\newcommand{\INAFRname}{INAF, Istituto di Astrofisica e Planetologia Spaziali, Via del Fosso del Cavaliere 100, 00133, Roma, Italy}

\newcommand{\INAFname}{INAF, Osservatorio Astronomico di Brera, via Bianchi 46, 23807 Merate (LC), Italy}

\newcommand{\INAFMname}{INAF, Istituto di Astrofisica Spaziale e Fisica Cosmica di Milano, Via Alfonso Corti 12, I—20133 Milano, Italy}

\newcommand{\INRname}{Institute  for Nuclear Research of the Russian Academy of Sciences, Moscow, Russian Federation}
\newcommand{\USiegenname}{Center for Particle Physics Siegen, Universit\"at Siegen, Siegen, Germany}
\newcommand{\CEAIrfuname}{IRFU, CEA, Universit\'e Paris-Saclay, F-91191 Gif-sur-Yvette, France}
\newcommand{\CEAListname}{Universit\' Paris-Saclay, CEA, List, Laboratoire National Henri Becquerel (LNE-LNHB), CEA‑Saclay, 91120 Palaiseau, France}
\newcommand{\MIPTname}{Moscow Institute of Physics and Technology, Dolgoprudny, Russian Federation}

\newcommand{\SOLEILname}{Synchrotron SOLEIL, 91192 Gif-sur-Yvette, France }
\newcommand{\MPPname}{Max-Planck-Institut f\"{u}r Physik, Boltzmannstr. 8, 85748 Garching, Germany}
\newcommand{\MPPHname}{Max Planck Institute for Nuclear Physics, Saupfercheckweg 1, 69117 Heidelberg, Germany}

\newcommand{\TUMname}{Technische Universit\"{a}t M\"{u}nchen, James-Franck-Str. 1, 85748 Garching, Germany}
\newcommand{\INFNname}{INFN, Istituto Nazionale di Fisica Nucleare, Sezione di Padova, Via F.\ Marzolo 8, 35131 Padova, Italy}

\newcommand{\UPADOVAname}{Dipartimento di Fisica e Astronomia ``Galileo Galilei'', Universit\`a degli Studi di Padova, Via F.\ Marzolo 8, 35131 Padova, Italy}
\newcommand{\CARTAGENAname}{Universidad Politécnica de Cartagena (UPCT)
Plaza del Hospital, 1. 30202 - Cartagena, Spain.}
\newcommand{\TUDOname}{Fakult\"{a}t f\"{u}r Physik, TU Dortmund, Otto-Hahn-Str. 4, Dortmund D-44221, Germany}
\newcommand{\UHHname}{Institute for Experimental Physics, University of Hamburg, Hamburg, 22761, Germany}
\newcommand{\SPAUKname}{School of Physics and Astronomy, University of Birmingham, Birmingham, B15 2TT, UK}
\newcommand{\INFNFname}{Galileo Galilei Institute for theoretical physics, Centro
Nazionale INFN di Studi Avanzati, 
Largo Enrico Fermi 2, I-50125, Firenze, Italy}
\newcommand{\BARRYname}{Physical Sciences, Barry University, 11300 NE 2nd Ave., Miami Shores, FL 33161, U.S.A.}

\newcommand{\ICCUB}{1}
\newcommand{\UNIZAR}{2}
\newcommand{\UHH}{3}
\newcommand{\INAF}{4}
\newcommand{\CEAIrfu}{5}
\newcommand{\UBonn}{6}
\newcommand{\CEFCA}{7}
\newcommand{\MIPT}{8}
\newcommand{\INR}{9}
\newcommand{\MPP}{10}
\newcommand{\INAFR}{11}
\newcommand{\PNPI}{12}
\newcommand{\CARTAGENA}{13}
\newcommand{\USiegen}{14}
\newcommand{\DESY}{15}
\newcommand{\CERN}{16}
\newcommand{\INAFB}{17}
\newcommand{\INFNB}{18}
\newcommand{\INAFM}{19}
\newcommand{\VATICAN}{20}
\newcommand{\UHEI}{21}
\newcommand{\BARRY}{22}
\newcommand{\SOLEIL}{23}
\newcommand{\UPADOVA}{24}
\newcommand{\INFN}{25}
\newcommand{\RBI}{26}
\newcommand{\Columbia}{27}
\newcommand{\CEAList}{28}
\newcommand{\MPPH}{29}
\newcommand{\TUM}{30}
\newcommand{\ICREA}{31}
\newcommand{\SPAUK}{32}
\newcommand{\INFNF}{33}
\newcommand{\LLNL}{34}
\newcommand{\Mainz}{35}
\newcommand{\TUDO}{36}
\newcommand{\INAFT}{37}
\newcommand{\INFNR}{38}
\newcommand{\UCT}{39}
\newcommand{\AIMS}{40}

\affiliation[\ICCUB]{\ICCUBname}
\affiliation[\UNIZAR]{\UNIZARname}
\affiliation[\UHH]{\UHHname}
\affiliation[\INAF]{\INAFname}
\affiliation[\CEAIrfu]{\CEAIrfuname}
\affiliation[\UBonn]{\UBonnname}
\affiliation[\CEFCA]{\CEFCAname}
\affiliation[\MIPT]{\MIPTname}
\affiliation[\INR]{\INRname}
\affiliation[\MPP]{\MPPname}
\affiliation[\INAFR]{\INAFRname}
\affiliation[\PNPI]{\PNPIname}
\affiliation[\CARTAGENA]{\CARTAGENAname}
\affiliation[\USiegen]{\USiegenname}
\affiliation[\DESY]{\DESYname}
\affiliation[\CERN]{\CERNname}
\affiliation[\INAFB]{\INAFBname}
\affiliation[\INFNB]{\INFNBname}
\affiliation[\INAFM]{\INAFMname}
\affiliation[\VATICAN]{\VATICANname}
\affiliation[\UHEI]{\UHEIname}
\affiliation[\BARRY]{\BARRYname}
\affiliation[\SOLEIL]{\SOLEILname}
\affiliation[\UPADOVA]{\UPADOVAname}
\affiliation[\INFN]{\INFNname}
\affiliation[\RBI]{\RBIname}
\affiliation[\Columbia]{\Columbianame}
\affiliation[\CEAList]{\CEAListname}
\affiliation[\MPPH]{\MPPHname}
\affiliation[\TUM]{\TUMname}
\affiliation[\ICREA]{\ICREAname}
\affiliation[\SPAUK]{\SPAUKname}
\affiliation[\INFNF]{\INFNFname}
\affiliation[\LLNL]{\LLNLname}
\affiliation[\Mainz]{\Mainzname}
\affiliation[\TUDO]{\TUDOname}
\affiliation[\INAFT]{\INAFTname}
\affiliation[\INFNR]{\INFNRname}
\affiliation[\UCT]{\UCTname}
\affiliation[\AIMS]{\AIMSname}

\newcommand{\IAXOAuthorList}{
S.~Ahyoune$^{\ICCUB}$,
K.~Altenm\"uller$^{\UNIZAR}$,
I.~Antol\'in$^{\UNIZAR}$$^{,}$$^{\UHH}$,
S.~Basso$^{\INAF}$,
P.~Brun$^{\CEAIrfu}$,
F.~R.~Cand\'on$^{\UNIZAR}$,
J.~F.~Castel$^{\UNIZAR}$,
S.~Cebri\'an$^{\UNIZAR}$,
D.~Chouhan$^{\UBonn}$,
R.~Della~Ceca$^{\INAF}$
% ...
}

\author{\IAXOAuthorList}

%% file: sections/01_introduction.tex
The search for axions and axion-like particles (ALPs) remains one of the most appealing and compelling experimental goals in modern physics, as an axion discovery would not only solve fundamental issues in particle physics but also have a profound impact on astrophysics and cosmology. 
Arguably, the most well-known and motivated ALP is the QCD axion, commonly referred to simply as ``the axion".
Axions were originally introduced to solve the strong CP problem in Quantum Chromodynamics (QCD) via the Peccei-Quinn mechanism\,\cite{Peccei:1977ur,Peccei:1977hh,Kim:2008hd}. 
However, it was soon shown that, due to non-thermal production mechanisms, they may contribute a substantial fraction of the dark matter in the universe\,\cite{Preskill:1982cy,Abbott:1982af,Dine:1982ah,Turner:1983he,Turner:1985si,1201.5902}.
More general ALPs, whose existence is not related to the strong CP problem, share much of the axion phenomenology and can be detected with similar strategies\,\cite{PaolaArias2012,PhysRevD.85.035018,Melcon_2018,PhysRevLett.117.141801}. 
They arise naturally in many extensions of the Standard Model, including string theory. In fact, they appear to be a natural consequence of the compactification of extra dimensions (see, e.g., Ref.~\cite{Jaeckel:2010ni}). 
Unlike axions, where a specific relationship between coupling and mass is expected, no such constraint exists for ALPs. This significantly broadens the viable parameter space for their detection, and expands the horizons on the search for these hypothetical particles (see, e.g., Refs.~\cite{Antel:2023hkf,Jaeckel:2010ni,Arias:2012az,Daido_2017}). 
While ALPs are generally pseudoscalar particles, scalar fields are also of interest, especially the so-called Chameleon particles which are additional targets of BabyIAXO and can be detected in ALP type experiments \cite{Steffen:2010ze,Upadhye:2009iv,GammeV:2008cqp,OShea:2024jjw}. Chameleon particles are a result of chameleon gravity~\cite{Khoury:2003rn, Khoury:2003aq}, an alternative theory to General Relativity that potentially explains the observed accelerated expansion of the universe~\cite{Brax:2004qh,Brax:2004px} while remaining consistent with all known tests of General Relativity to date.

Building on previous developments in helioscope experiments such as CAST\,\cite{ZIOUTAS1999480,PhysRevLett.94.121301,Andriamonje_2007,Arik_2009,PhysRevLett.112.091302,Anastassopoulos2017}, the International Axion Observatory (IAXO)\,\cite{Irastorza_2011,Armengaud_2014} aims to achieve unprecedented sensitivity in detecting solar axions and ALPs using the axion helioscope technique\,\cite{Sikivie:1983ip}.
Its pathfinder setup, BabyIAXO, serves as a crucial intermediate step in this ambitious endeavor, functioning both as a technological prototype and an independent physics experiment\,\cite{Abeln2021}. BabyIAXO's primary goal is to test and validate key subsystems—such as the magnet, optics, and detectors—all of which are essential for the full-scale IAXO experiment. Additionally, BabyIAXO has the potential to explore new regions of ALP parameter space, including significant portions of the QCD axion band\,\cite{Armengaud_2019,OShea_2024}. Utilizing spectral information, IAXO may also provide valuable insights into the axion couplings to photons, electrons, and nucleons\,\cite{Jaeckel:2018mbn,DiLuzio2022}.

\section{BabyIAXO}

\begin{figure}[hbt!]
	\begin{center}
        \includegraphics[width=0.98\linewidth]{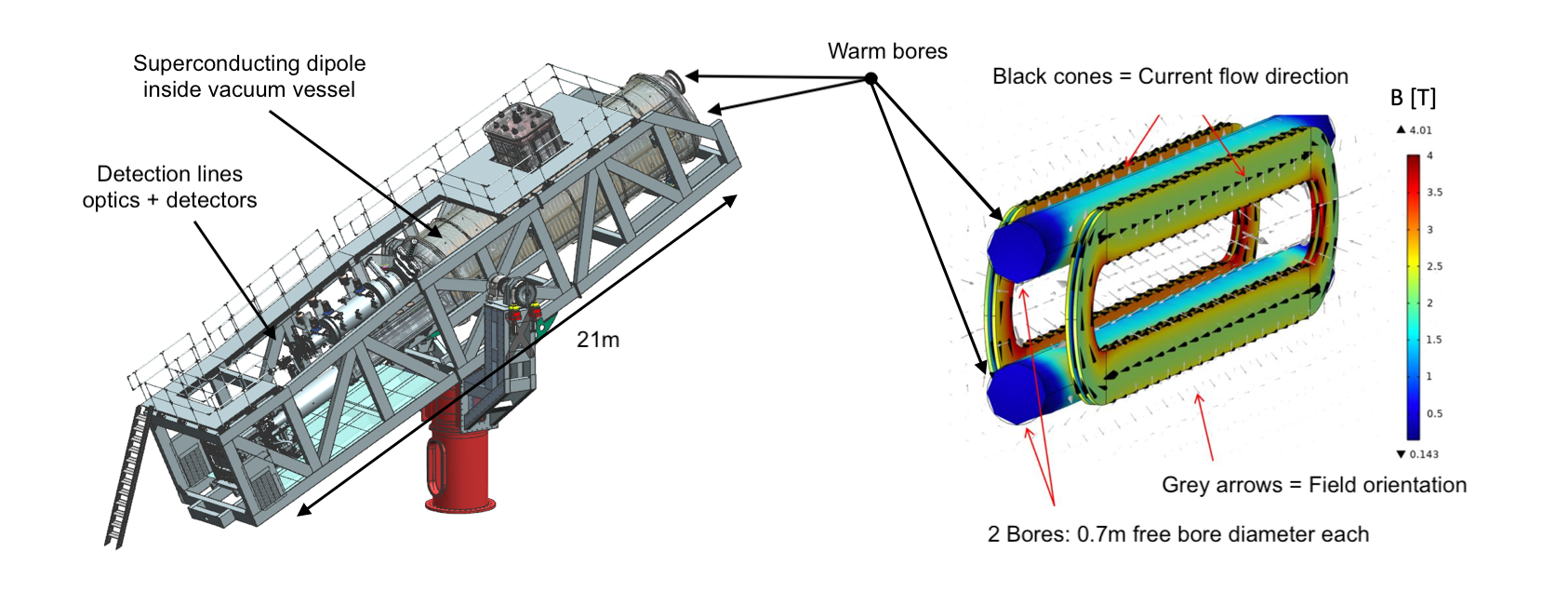}
		\caption{On the \textit{left}, the latest version design of the BabyIAXO moving structure hosting the different helioscope components (currently under revision). A vacuum vessel will host a superconducting dipole magnet providing an intense magnetic field where axion-photon conversion takes place. On the \textit{right}, the magnetic field configuration generated by the superconducting dipole.}
		\label{fig:BabyIAXO}
	\end{center}
\end{figure}

BabyIAXO's design (see~Figure~\ref{fig:BabyIAXO}) has been optimized to probe previously unexplored regions of the ALP and axion parameter space, targeting areas strongly motivated by both cosmology and astrophysics \cite{IAXO:2019mpb,Abeln2021}. The helioscope's drive system has been designed to assure proper alignment with the sun line of sight and provide an effective tracking time of at least 12\,hours per day. Axions generated in the sun's core travel all the way towards the helioscope's magnetic field region, where they can be converted into photons. These photons are then focused by the X-ray optics placed behind the magnet into a small detection area, or spot. The experiment enhances its sensitivity through the accumulation of long exposure times, with data collection spanning several years.

The experimental program of BabyIAXO includes two data-taking campaigns: the vacuum phase and the gas phase. In the vacuum phase, any gas present in the magnetic field region will be evacuated, optimizing the sensitivity to the low-mass axion range. The gas phase introduces a light atomic buffer gas into the magnetic field region, which enhances sensitivity to higher-mass axions by adjusting the axion-photon resonance condition. Both phases are essential to extend the sensitivity of BabyIAXO over a wide range of axion masses favored by theoretical QCD axion models.

BabyIAXO benefits from a dedicated magnet design specifically developed for solar axion searches, substantially increasing its figure of merit (FOM) compared to previous searches due to the large magnet aperture. One of the detection lines will use a spare XMM-Newton optics while the second line will host a custom-designed X-ray optics design. An X-ray detector placed at the focusing region, and optimized with ultra-low background rare event techniques\,\cite{ALTENMULLER2023167913}, will magnify the signal-to-noise ratio further. Unlike in accelerator physics, the homogeneity of the magnetic field is not a strict requirement in axion helioscope physics. Therefore, it is necessary to include the BabyIAXO inhomogeneous magnetic field profile in our calculations and conduct a more refined computational study to validate the expected axion signal. In addition, we have to convolve the solar axion flux distribution with the larger magnet aperture.

We have developed an accurate model of BabyIAXO's response to solar axions through ray-tracing. A rigorous software implementation of the helioscope components, including the magnet, X-ray optics, and detectors, allows for detailed analysis of signal sensitivity and to assess the impact of future potential upgrades on the experiment performance. By simulating the conversion of solar axions into X-rays within the magnetic field of BabyIAXO and tracing these photons through advanced optics and detection systems, we provide a precise estimate of the instrument’s sensitivity to the axion-photon coupling,~$g_{a\gamma}$.

The following sections offer a detailed overview of the ray-tracing model, software tools, and sensitivity analysis, showcasing BabyIAXO’s potential to make significant advancements in the search for axions and to address fundamental questions in astrophysics and cosmology.

Our software implementation accurately reproduces the experimental conditions of these campaigns, enabling us to model axion-photon conversion and photon propagation under both vacuum and gas-filled conditions. The progress presented in this work is essential to evaluate the impact a large-aperture inhomogeneous magnetic field would have on the experimental sensitivity of BabyIAXO, which had previously only been assessed under the constant-field approximation.

%% file: sections/02_components.tex
\section{Software implementation of helioscope components}
\label{sc:components}

The different axion helioscope components required to calculate the apparatus response have been implemented inside the \textit{REST-for-Physics} framework\,\cite{ALTENMULLER2022108281} as metadata descriptors, consisting of C++ data members that define the behavior of a particular class, and that can be integrated inside the \textit{REST-for-Physics} framework I/O scheme, based on \textit{ROOT}\,\cite{ANTCHEVA20092499}. The new classes have been contributed to a dedicated \textit{REST-for-Physics} library for axion physics---\textit{axionlib}\,\cite{jgl}---which can be optionally compiled as a module into the \textit{REST-for-Physics} framework. The classes have been designed to describe and provide access to the main helioscope components: the solar axion flux (\emph{SolarFlux}\footnote{The class naming convention inside \textit{REST-for-Physics} uses a common root, \emph{TRest}, for the class definitions, and an extended common root, or prefix, to designate the belonging to a particular library, such as the \emph{TRestAxion} prefix for classes belonging to the \textit{REST-for-Physics} axion library. Those prefixes will be omitted in the main text while an enhanced text will indicate the direct relation with the code repository class.}), the magnetic field map (\emph{MagneticField}), the optics response (\emph{Optics}), and the X-ray windows transmission (\emph{XrayWindow}). The software implementation of those components has been carefully designed to allow future modifications, e.g.\ adding new solar axion flux components, loading a different field map configuration, integrating a new focusing device, or creating different X-ray windows setups.

This section details the specific components used in producing ray-tracing MC data for this work and explains how the software metadata describes the main features of each BabyIAXO helioscope component.

\input{sections/02_components_flux}

\input{sections/02_components_magnet}

\input{sections/02_components_optics}

\input{sections/02_components_windows}

%% file: sections/02_components_flux.tex
\subsection{Solar axion flux}\label{sc:flux}

The sun produces ALPs through a variety of mechanisms, depending on their couplings to photons ($g_{a\gamma}$), electrons ($g_{ae}$) and nucleons.

The most important production mechanisms associated with $g_{a\gamma}$ are the Primakoff process, originally investigated in Refs.~\cite{Dicus:1979ch,Fukugita:1982ep,Fukugita:1982gn,Raffelt:1985nk}, and plasmon conversion~\cite{Raffelt:1987np,Caputo:2020quz,OHare:2020wum,Guarini:2020hps}.
Primakoff production converts photons into ALPs inside the electromagnetic fields of electrons and ions in the plasma, $\gamma + Ze \to Ze + a$, while the plasmon conversion is most relevant in the sun's macroscopic magnetic fields.
Another important difference is that the theoretical uncertainties on the Primakoff flux are at the percent level~\cite{Hoof:2021mld}, while the structure and intensity of the sun's macroscopic magnetic field is not well known and hence subject to much larger uncertainties.

For $g_{ae}$, there are a number of processes to consider, namely the ``ABC processes'' of atomic recombination and de-excitation, axion bremsstrahlung in electron-electron ($e$-$e$) and electron-ion collisions, and Compton scattering~\cite{Mikaelian:1978jg,Zhitnitsky:1979cn,Fukugita:1982ep,Fukugita:1982gn,Krauss:1984gm,Raffelt:1985nk,Dimopoulos:1986kc,Dimopoulos:1986mi,Pospelov:2008jk,Redondo:2013wwa}.

Axions coupled to nucleons can be produced via nuclear fusion and decays, as well as thermal excitation and subsequent de-excitation of the nuclei of stable isotopes~(see, e.g., Refs.~\cite{Moriyama:1995bz,Krcmar:1998xn,DiLuzio2022}).
While there is a large number of possible ALP production channels from nuclear reactions inside the sun, the corresponding expected axion fluxes are comparably small.
The most relevant examples include proton-deuterium fusion ($p + \ce{^2H} \to \ce{^3He} + a$; \SI{5.5}{\MeV}), lithium de-excitation ($\ce{^7Li^*} \to \ce{^7Li} + a$; \SI{0.478}{\MeV}), and
\ce{^{57}Fe} de-excitation ($\ce{^{57}Fe^*} \to \ce{^{57}Fe} + a$; \SI{14.4}{\keV}).
Note that the axion-nucleon coupling is an effective quantity, which has to be specified for each of nuclear process since it depends on a combination of couplings to protons and neutrons (see, e.g., Refs.~\cite{DiLuzio2022,Carenza:2024ehj}).

In this work, we focus on the Primakoff and ABC processes.
The spectral solar axion flux can then be written as
\begin{equation}
\frac{\dd N_a}{\dd t \, \dd\omega}
= \left( \frac{\gagamma}{\gagamma^\star} \right)^2 n_{a\gamma}^\star(\omega) + \left(\frac{\gae}{\gae^\star}\right)^2 n_{ae}^\star(\omega)
 \, , \label{eq:solar_flux_1}
\end{equation}
where $n_{a\gamma}^\star$ and $n_{ae}^\star$ respectively are the flux contributions from the Primakoff and ABC processes at some reference values of the couplings, $\gagamma^\star$ and $\gae^\star$.

\begin{figure}[htb]
	\begin{center}
    \includegraphics[width=0.98\linewidth]{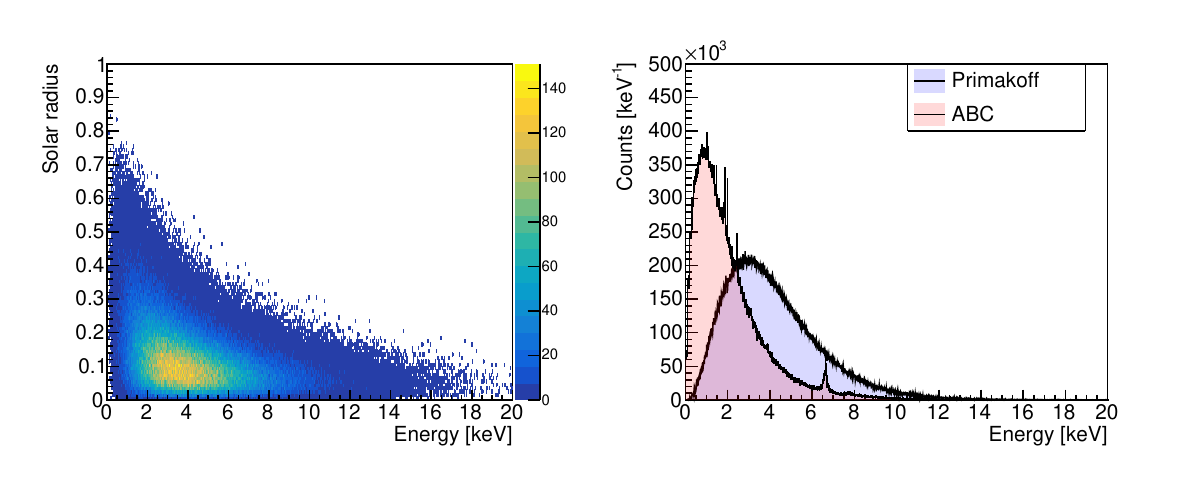}
	\caption{Binned spectral and radial solar axion flux from a Monte Carlo simulation with \num{e6}~events for the B16-AGSS09 solar model. On the \textit{left}, the two-dimensional distribution shows the correlation between the normalized solar radius and the axion energy for the Primakoff flux component. On the \textit{right}, the one-dimensional distribution shows the dependence on the axion energy alone.}
	\label{fig:solarFlux}
	\end{center}
\end{figure}

The \textit{REST-for-Physics} axion library ---\textit{axionlib}~\cite{jgl}--- implements the spectral solar axion flux components in \cref{eq:solar_flux_1} using tabulated values computed with the external SolarAxionFlux library~\cite{Hoof:2021mld} for the B16-AGSS09 solar model~\cite{Vinyoles:2016djt}.\footnote{We acknowledge contributions from Lennert J.\ Thormaehlen in preparing these tables.}
In particular, the \emph{SolarQCDFlux} metadata class uses these tabulated spectral fluxes for Monte Carlo (MC) event generation.
In addition to the spectral dependence, we can also consider the radial dependence of the flux components -- i.e.\ as a function of the radius~$\rho$ on the solar disc.
We include this option in \cref{fig:solarFlux}, where we show the radial and spectral fluxes (left panel) in addition to the spectral dependence alone (right panel) as obtained from a MC simulation with \num{e6} events.
The $\rho$~dependence can be useful as it has been shown that it may be used to infer the solar temperature profile in case of an axion detection~\cite{Hoof:2023jol}.
Moreover, since the ABC flux is peaked at slightly lower energies than the Primakoff process, we may even use the spectral information alone to separately measure $\gagamma$ and $\gae$~\cite{Jaeckel:2018mbn}.

For the smooth components of the solar axion flux, we can fit analytical expressions to the numerical results using previously proposed fitting {formul\ae}~\cite{Andriamonje:2007ew,Barth:2013sma}.
For the B16-AGSS09 solar model~\cite{Vinyoles:2016djt}, $\gagamma^\star = \SI{e-10}{\GeV^{-1}}$, and $\gae^\star = \num{e-13}$, we find the following numerical results values, updating parts of the literature where inconsistent or outdated solar models have been used:\footnote{Recently, new solar models have been put forward based on updated abundances, which claim to resolve the ``solar abundance problem''~\cite{Magg:2022rxb}. While there is at least some debate about this claim~\cite{Buldgen:2022nso,Buldgen:2024aak}, any future data analysis should update the tables to future state-of-the-art solar models.}
\begin{align}
\frac{n_{a\gamma}^\star(\omega)}{\si{\cm^{-2} \s^{-1} \keV^{-1}}} &\approx \num{5.84e10} \, \omega^{2.51} \, \mathrm{e}^{-0.850 \, \omega} \, , \\
    \frac{n_{ae}^\star(\omega)}{\si{\cm^{-2} \s^{-1} \keV^{-1}}} &\approx \num{4.89e9} \frac{\omega \, \mathrm{e}^{-0.819 \, \omega}}{1 + 0.941 \, \omega^{1.12}} + \num{6.78e6} \, \omega^{2.98} \, \mathrm{e}^{-0.786 \,\omega} + \dots \, , \label{eq:flux_approx_abc}
\end{align}
where $\omega$ is in units of \si{\keV} and where the first term in \cref{eq:flux_approx_abc} corresponds to axion production from $e$-$e$ and $e$-ion bremsstrahlung, while the second one approximates the ALP emission from a Compton-like process.
Note that, for the ABC flux, two independent tables are provided to the MC generator: one containing the smooth, slowly varying spectral component, and another with a more refined table (10\,eV energy resolution) that describes the sharp spectral features originating from monochromatic lines generated by atomic transitions and other physical processes taking place in the solar medium. Both tables are combined in the MC simulation to reproduce the ABC flux shown in \cref{fig:solarFlux}.
The relative intensity of Primakoff and ABC fluxes depends on the specific axion model. In general, if $\gagamma \gtrsim \SI{8.7e-11}{\GeV^{-1}} (\gae/\num{e-13})$, Primakoff production will dominate the solar axion flux~\cite{Hoof:2021mld}.

%% file: sections/02_components_magnet.tex
\subsection{Magnetic field}\label{sec:magnet}

The BabyIAXO magnet features a common-coil configuration, generating a transverse magnetic field of approximately 2\,T across two free bores, each with a diameter of 70\,cm. The superconducting cold mass is constructed from aluminum-stabilized Nb-Ti/Cu cables, arranged in pancake windings, and operates at a current of 6\,kA. The cryogenic cooling system is innovative, utilizing cryo-coolers to cool helium gas to 50\,K and 25\,K for thermal shield cooling, and to locally condense liquid helium at 4.5\,K. Currently, the magnet design is being prepared for industrial fabrication, with parallel efforts underway to procure the required aluminum-stabilized conductor. Unlike its predecessors, the BabyIAXO magnet is designed with large-aperture bores in which the magnetic field is non-homogeneous~\cite{CAST:2024eil,OHTA201273}.

\subsubsection{Axion probability and field map description}

A proper convolution of the axion solar flux with the inhomogeneous magnetic field profile and subsequent helioscope components, such as optics and window interfaces, may benefit from ray-tracing capabilities that require the evaluation of the transverse field component along the particle track. This involves a corresponding numerical integration of the field equations for axion motion propagating in a magnetic medium\,\cite{PhysRevD.39.2089,doi:10.1080/00107514.2011.563516}, which reduces to the computation of the probability of an axion to convert to an X-ray photon, given by the following expression:

\begin{equation}\label{eq:probability}
P_{a\gamma} = \frac{1}{4} g^2_{a\gamma} \mbox{exp}\left(-\Gamma L\right) \left| \int_L \mbox{exp}(\Gamma l/2 + iql) B_\perp (l) dl \right|^2
\end{equation}

\noindent where $l$ is the parameterization length along the particle trajectory, $\Gamma$ is the X-ray attenuation length in the medium, and $q=(m^2_a-m^2_\gamma)/2E_a$ is the corresponding momentum transfer. Here, $m_\gamma$ is determined by the medium's plasma frequency, $\omega^ 2_p=m^2_\gamma=4\pi r_o (N_A / A m_u) \rho f_1$, where $r_o$ is the classical electron radius, $N_A$ is the Avogadro's number, $A$ is the atomic mass number of the medium, $m_u$ is the atomic mass unit, $\rho$ the gas medium density, and $f_1$ is the atomic scattering form factor.

The calculations needed to determine $\Gamma$ and $f_1$ as functions of energy have been implemented inside the \textit{REST-for-Physics} \emph{BufferGas} axion metadata class, enabling the use of any gas mixture whose atomic properties can be extracted from the NIST database. For BabyIAXO we use a pure helium gas at different densities, while we can set $\Gamma=0$ and $m_\gamma=0$ in equation\,(\ref{eq:probability}) for the vacuum case. 

\begin{figure}[ht!]
	\begin{center}
		\includegraphics[width=0.98\linewidth]{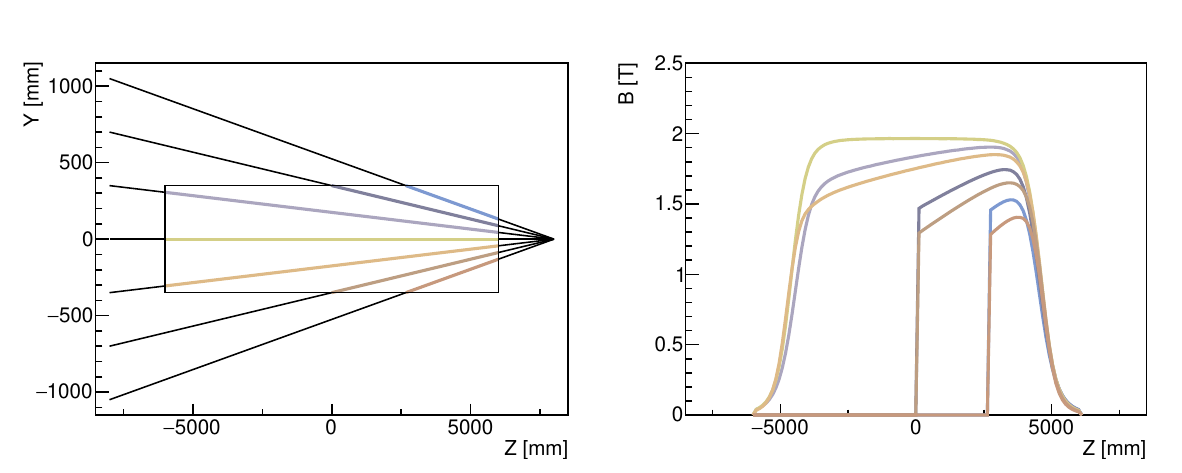}
		\caption{On the \textit{left}, a bounding box illustrates the cylindrical magnet projection on the YZ-plane where we observe test particles traversing the magnetic region with different trajectories. On the \textit{right}, the transverse field component obtained along the path for each particle using the same color legend as on the left plot.}
		\label{fig:magnetTracks}
	\end{center}
\end{figure}

The numerical integration of equation\,(\ref{eq:probability}) has been implemented inside the \emph{Field} class of \textit{axionlib}. The \emph{Field} class utilizes the \emph{BufferGas} construction, and the \emph{MagneticField} vectorial map description. The \emph{MagneticField} class not only provides interpolation routines to evaluate the field at any point but also serves to reconstruct the magnetic field profile along a chosen track (see Figure\,\ref{fig:magnetTracks}). We assume the trajectories to be straight lines, as prior studies have shown that the bending of the axion-photon wave due to magnetic field inhomogeneities is negligible in our case~\cite{Redondo:2010js}.

The field map used in this work has been generated using a Finite Elements Method (FEM) COMSOL simulation of the latest magnet design and can be found in the \textit{REST-for-Physics} axion library repository. The original input field map resolution is $\Delta x \times \Delta y \times \Delta z$ = 10$\times$10$\times$50\,mm$^3$, where $z$ is aligned with the magnet axis. The field map description extends over a region from $z = \SI{-10}{\m}$ to $z = \SI{10}{\m}$ along this axis, although the effective length where the field intensity is strong enough is only about \SI{10}{\m}. In the other dimensions, the map covers a circular area of radius 35\,cm in the $xy$-plane. The field map can be used to determine the magnet FOM (MFOM) by computing a three-dimensional integral of the transverse magnetic field using the expression:

\begin{equation}
    \mbox{MFOM (3-D)} = \int_A \left( \int_L B_\perp(x,y,z) dz \right)^2 dx\,dy \, ,
\end{equation}

\noindent where the integral along the length \emph{L} runs over all possible trajectories parallel to $z$ in the magnet aperture $A$. The calculation of this integral for BabyIAXO yields 265\,T$^2$m$^3$, which is equivalent to an average field of $B_\perp = \SI{2}{\tesla}$ for an effective length of $L = \SI{9.3}{\m}$.

\subsubsection{Numerical integration}\label{sec:integration}

The relation in \cref{eq:probability} is integrated within the \emph{Field} class using the GSL libraries\,\cite{gough2009gnu}. The QAG adaptive integration method is used when the momentum transfer is $q=0$, while the QAWO adaptive integration method for oscillatory functions is used when $q\neq0$. The GSL integration routines return both the integral result and its associated error. Several factors, such as the boundaries of the field map definition, the resolution of field grid, and the tolerance of the integration method, can affect the integral result and its error.

\begin{figure}[ht!]
	\begin{center}
		\includegraphics[width=0.66\linewidth]{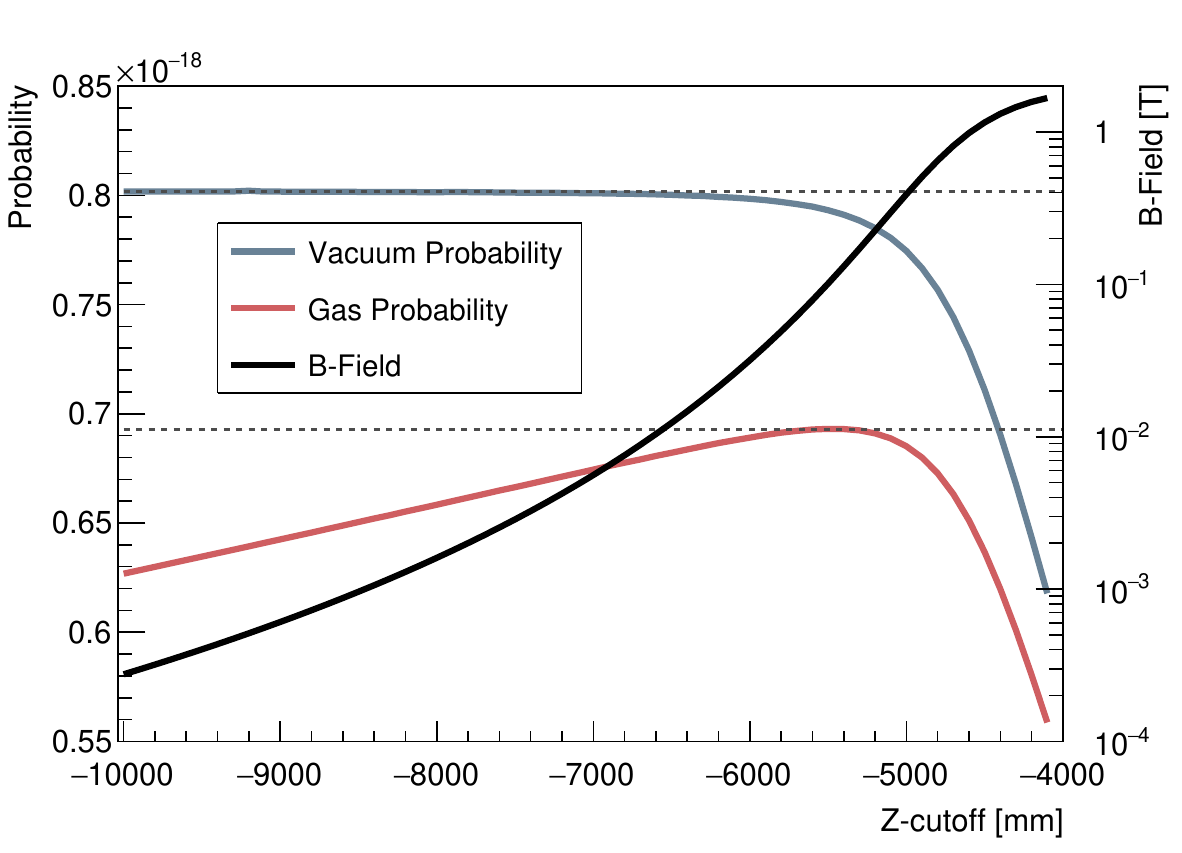}
		\caption{The magnetic field profile (logarithmic scale) along the $z$-axis for the first 6~meters of the field map definition is shown, together with the axion photon probability calculated using different integration limits over the interval $(-Z_\text{cutoff},\,Z_\text{cutoff})$. The calculation assumes a photon energy of 4.2\,keV. The buffer gas helium density is $\rho$=0.327\,mg/cm$^3$, corresponding to an axion mass of 0.368\,eV. In both cases the axion mass used in the calculation was in resonance with the medium. For reference, using a constant L=9.3\,m and B=2\,T the axion-photon conversion probability in vacuum is higher, at 8.48$\times$10$^{-19}$.}

		\label{fig:fieldCutOff}
	\end{center}
\end{figure}

A fundamental question in the field numerical integration is the choice of the field map boundaries. Indeed, as the field boundaries along the $z$-axis are approached, the field intensity rapidly decreases. In \cref{fig:fieldCutOff} we observe how the field---plotted on a logarithmic scale---drops in intensity at around $z = -5\,\text{m}$ as we move further from the center of the field map, placed at $z = 0\,\text{m}$.
re\,\ref{fig:fieldCutOff} illustrates how the choice of integration limits affects the integral's result. In vacuum, the axion-photon conversion probability approaches its maximum around $z = -6\,\text{m}$. In contrast, for a buffer gas filling the magnetic field volume, a clear maximum probability is observed around $z = -5.5\,\text{m}$. In this latter case, extending the field map boundaries further results in a decrease in axion-photon probability due to the additional gas photon absorption in the end region. This graph suggests then the optimal position for a differential window separating the buffer gas medium from a vacuum region where photons would not undergo gas absorption.

Furthermore, we must consider the impact of the field map spatial resolution on the probability calculation. Table\,\ref{tab:grid} summarizes the results obtained for different cell sizes. In resonance conditions ($\Delta m$=0\,meV), the impact of cell resolution on the probability is acceptable up to cell sizes of  40$\times$40$\times$200\,mm$^3$. However, the impact is more pronounced when the integral is calculated outside the resonance, at $\Delta$m=10\,meV. Here, deviations become noticeable starting at 20$\times$20$\times$100\,mm$^3$ and become unacceptable for cell sizes equal to and greater than 40$\times$40$\times$200\,mm$^3$.

\begin{table}[htb]
    \centering
    \begin{tabular}{c|cccc}
        \multirow{2}{*}{$\Delta x \times \Delta y \times \Delta z$ [mm$^3$]} & \multicolumn{3}{c}{Probability} \\
        & $\Delta m$=0\,meV & 1\,meV & 10\,meV  \\
        \hline
        10$\times$10$\times$50 & 6.273$\pm$0.003 & 1.109$\pm$0.015 & 3.677$\pm$0.029  \\
        20$\times$20$\times$50 & 6.273$\pm$0.002 & 1.109$\pm$0.015 & 3.663$\pm$0.037  \\
        20$\times$20$\times$100 & 6.272$\pm$0.022 & 1.109$\pm$0.015 & 3.579$\pm$0.036  \\
        40$\times$40$\times$100 & 6.269$\pm$0.015 & 1.105$\pm$0.017 & 3.628$\pm$0.025  \\
        40$\times$40$\times$200 & 6.268$\pm$0.012 & 1.103$\pm$0.016 & 3.335$\pm$0.045  \\
        40$\times$40$\times$400 & 6.304$\pm$0.041 & 1.099$\pm$0.020 & 2.200$\pm$0.029  \\
        \hline
         & $\times$10$^{-19}$ & $\times$10$^{-19}$ & $\times$10$^{-23}$ \\

    \end{tabular}
    \caption{Axion-photon conversion probability in gas for different field map granularities, and different axion mass offsets, $\Delta m$, from the resonance mass. Calculated using the same conditions to those given in Figure\,\ref{fig:fieldCutOff}. The integration method's tolerance was chosen equal to 0.01.}
    \label{tab:grid}
\end{table}

The tolerance of the numerical integration method also affects the result. Lower tolerances yield lower errors but may cause the integral computation diverge, leading the method to fail to return a value. This issue is observed in Table~\ref{tab:tolerance}, where convergence problems occur for larger off-resonance cases.
For example, for $\Delta m=100\,\text{meV}$, no response is obtained for lower tolerances, while at higher tolerances, the error is compatible with zero, making it impossible to obtain an accurate probability value. Despite this, the estimated value does not impact the final sensitivity, as the off-resonance axion signal at $\Delta m$=100\,meV is more than 10 orders of magnitude below the probability obtained at the resonance.

\begin{table}[htb]
    \centering
    \begin{tabular}{c|cccc}
        \multirow{2}{*}{Tolerance} & \multicolumn{4}{c}{Probability} \\
 & $\Delta m = \SI{0}{\meV}$ & \SI{1}{\meV} & \SI{10}{\meV} & \SI{100}{\meV} \\
        \hline
        0.001 & 6.272$\pm$0.003 & 1.109$\pm$0.002 & N/A & N/A \\
        0.01 & 6.272$\pm$0.003 & 1.11$\pm$0.01 & 3.68$\pm$0.03 & N/A \\
        0.1 & 6.272$\pm$0.003 & 1.13$\pm$0.12 & 3.68$\pm$0.21 & 3.26$\pm$6.21 \\
        0.5 & 6.27$\pm$1.85 & 1.13$\pm$0.12 & 2.91$\pm$1.93 & 23$\pm$40 \\
        \hline
         & $\times$10$^{-19}$ & $\times$10$^{-19}$ & $\times$10$^{-23}$ & $\times$10$^{-30}$ \\

    \end{tabular}
    \caption{Axion-photon conversion probability in gas for different tolerances and various axion mass offsets, $\Delta m$, from the resonance mass,  calculated using the same conditions as those given in Figure\,\ref{fig:fieldCutOff}. The field map cell resolution chosen was 10$\times$10$\times$50\,mm$^3$.}
    \label{tab:tolerance}
\end{table}

%% file: sections/02_components_optics.tex
\subsection{X-ray optics}\label{sc:optics}

The BabyIAXO magnet will feature two \SI{70}{cm}-diameter ports, integrated into the helioscope as separate detection lines. Each port will be equipped with different optical instruments designed to focus X-ray photons, converted in the magnetic field, into a small spot or region where a low-background detector will be placed. By utilizing X-ray optics, the large magnet aperture can be fully exploited while reducing background noise by focusing on a small area of the detector only. This method allows for a more effective detector design, minimizing intrinsic background levels through rare event search techniques, such as using radiopure materials and implementing both passive and active shielding, which would be more challenging with larger detector setups.

\begin{figure}[t]
	\begin{center}
		\includegraphics[width=0.98\linewidth]{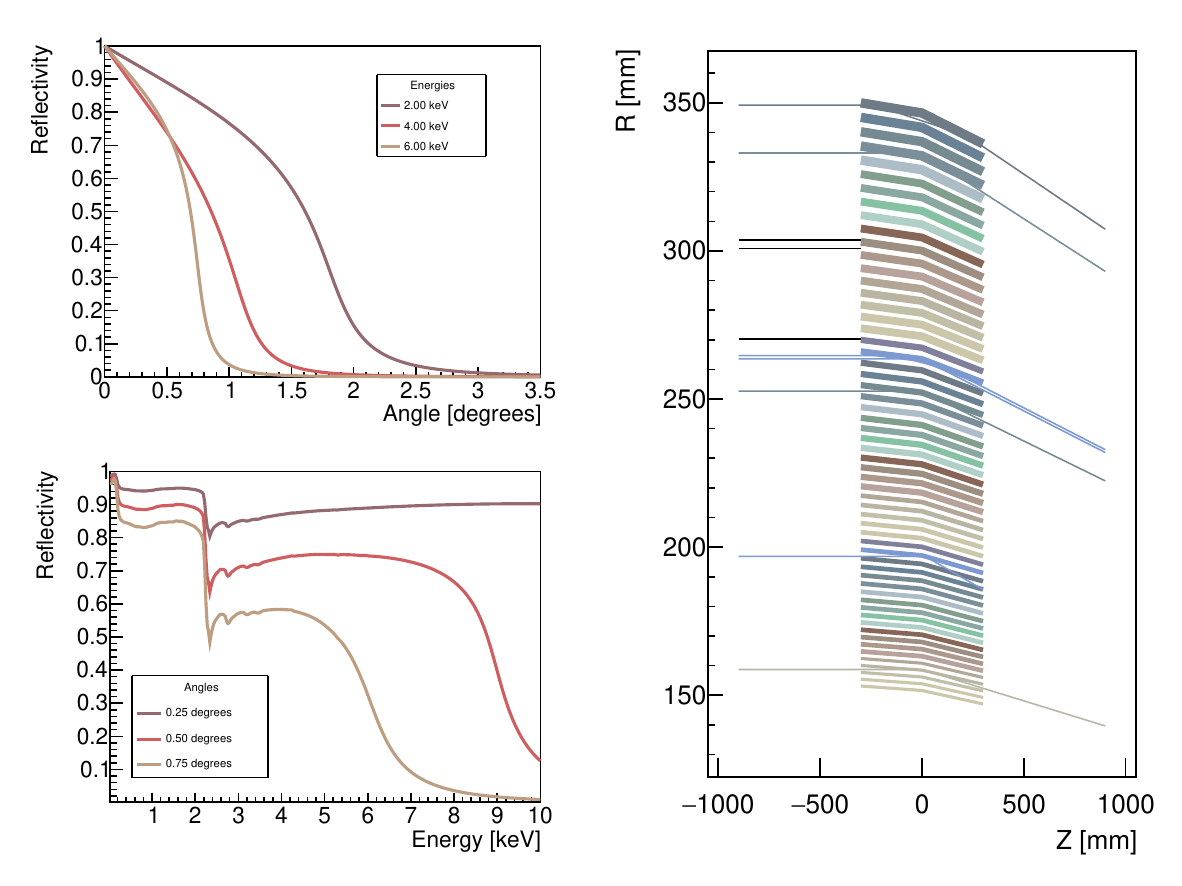}
		\caption{On the \textit{left}, the reflectivity extracted from the XMM mirror properties used on the optics implementation, as a function of the incidence angle (top), and as a function of energy (bottom). On the \textit{right}, a schematic illustrating the mirror shells in a section of the geometry. The mirror shells are represented with a line thickness proportional to the mirror shell thickness. Photons blocked in the first entrance mask are drawn in black. The photon tracks acquire the color of the mirror shell where they are reflecting. Only photons that experience two proper reflections and exit through the same inter-mirror aperture will continue their path towards the focal spot. For example, the second track from bottom enters the optics device but it is not properly reflected out.  }
		\label{fig:optics}
	\end{center}
\end{figure}

The baseline strategy involves covering one of the two magnet bores with custom-designed BabyIAXO optics, whose performance is comparable to a final IAXO optics.
The other bore will be equipped with a flight-spare module from the X-ray Multi-mirror Mission (XMM) Newton\,\cite{Jansen:2001bi}, using Wolter-I-type optics. While the custom optics is presently being implemented in the software library, the XMM optics has been already implemented inside the \emph{TrueWolterOptics} class and is available inside \textit{axionlib}, capable of simulating the reflections in a hyperbolic-parabolic mirror system. The reflection of photons in the mirrors is modeled using the optical properties of the materials employed in the device's construction (see Figure\,\ref{fig:optics}). These optical properties are encoded in the \emph{OpticsMirror} class, which utilizes the Henke database\,\cite{Henke:1993eda}. Specifically, the XMM mirror is made of a 250\,nm gold single layer on a nickel substrate.

The current optics software implementation does not include all the physics available in dedicated photon ray-tracing software for optical instruments; effects such as diffuse reflection, mirrors deformation, or as-built metrology have not been included, yet. However, integrating the reflection process of X-ray optics into the library will, for the first time, allow for the convolution of the axion-photon conversion probability with the optics system in an inhomogeneous magnetic field.
Additionally, common tools such as the \emph{PatternMask} from the \textit{REST-for-Physics} framework or the \emph{OpticsMirror} metadata class can be used together with the \emph{Optics} class interface to construct new optics. In general, constructing ray-tracing components requires defining regions of space where photons are not allowed to propagate; the \emph{PatternMask} class is used to define a spatial mask versatile enough to reproduce a realistic entrance pattern of the optics. An example for this usage is shown in Figure\,\ref{fig:optics}, where we observe that some photons are blocked due to the masks defined at different $z$-planes (see \cref{sec:appendixA} for more details). This class is also used to define masks in other ray-tracing components, such as the X-ray transmission windows described in \cref{sc:windows}.

\subsubsection{XMM optical efficiency estimation}

The ray-tracing results will be shown later in \cref{sc:sensitivity}. In order to compare those results with a simplified calculation of the signal, we will estimate the optical efficiency under certain assumptions. First, we must consider the mechanical structure that holds and aligns the XMM X-ray mirrors. This mechanical structure adds opacity to photon transmission. Specifically, we calculate the optical transparency considering both the spider arms structure (see~\cite{Abeln2021,Jansen:2001bi} for additional details) and the thickness of mirror shells, which block photons from entering the focusing device. Assuming that the thickness of the mirrors is much smaller than the radius of the mirror shells, $d_n\ll r_n$, the optical transmission is given by the following relation:

\begin{equation}
    T_o \simeq \frac{(1-\frac{n_\text{a}w_a}{2\pi})}{R_\text{out}^2}\left[(R_\text{out}^2-R_\text{in}^2)-{2\sum_n d_n r_n}\right]
\end{equation}

\noindent where $R_\text{out}=35$\,cm and $ R_\text{in}=15.3$\,cm are the radii of the outermost and innermost mirror shells, respectively, and $n_a=16$ and $w_a=\SI{2.29}{\degree}$ are the number of spider arms and its corresponding angular width.\footnote{These parameters can also be found on the optics geometry file, \emph{xmmTrueWolter.rml}, within the \textit{axionlib} data repository, library release 2.4.}

For photons that enter the focusing device, only those undergoing double reflection will be correctly imaged on the X-ray detector at the focal plane. To estimate the average response to the optics reflectivity, we need to convolve the squared reflectivity of the mirrors with the solar axion flux (given in section\,\ref{sc:flux}) as a function of energy using the following expression:

\begin{equation}\label{eq:Reff}
    R_\text{eff}^2 = \int_E\phi_a(E) R^2(E) \left/ \int_E\phi_a(E) \right.
\end{equation}

\noindent where $\phi_a(E)$ is the axion flux, and $R(E)$ is the reflectivity as a function of energy.

The overall optics efficiency can then be estimated by assuming an average incidence angle of \SI{0.2}{\degree} and considering the flux from Primakoff production. We obtain

\begin{equation}
    \epsilon_o = T_{o} \times R^2_\text{eff} =  0.57 \times 0.825 = 0.47.
\end{equation}

Another relevant optical parameter to consider in the sensitivity calculation is the spot size. A ray-tracing calculation using a realistic solar axion flux as input places the Half-Energy Width (HEW) of the focal spot at 7.83\,mm, while 90\% of the focused photons are contained within a diameter of 16.83\,mm. The spot size values will be used later on for estimating the background counts that must be taken into account in the sensitivity analysis. Those values, as obtained by the ray-tracer, have been found comparable to those obtained from experimental measurements of the HEW and the off-axis effective area.

%% file: sections/02_components_windows.tex
\subsection{Windows transmission}\label{sc:windows}

Photons converted in the magnetic field region travel through space, are focused by the optics, and ultimately reach the X-ray detector located at the focal plane. To maximize the detection efficiency of the helioscope and avoid photon absorption, it is important to minimize the distance over which photons  travel through the gas between the magnetic field and the detector. Therefore, when operating with a buffer gas inside the magnetic field region, it is desirable to use a differential window that maintains pressure within the region where the magnetic field is confined. This window must be sufficiently transparent in the X-rays energy range, absorbing far fewer photons than the buffer gas would if it filled the space between the magnet and the detector. In \cref{sec:magnet}, we identified the optimal position\footnote{Note that this optimal window position may vary slightly depending on the magnet gas density.} for this differential window (see Figure\,\ref{fig:fieldCutOff}).ecessary differential pressure window is required for gaseous particle detection technologies, such as Micromegas\,\cite{Giomataris:1995fq,Andriamonje:2010zz} or GridPix\,\cite{KRIEGER2017101}, which typically operate with a gas at atmospheric pressure or higher to increase their detection efficiency. The X-ray windows should be constructed using thin layers of materials with good photon transparency in the energy range between a few hundred eV to tens of keV. Additionally, the thin material layers used in the window construction must be robust enough to withstand the pressure difference. To enhance robustness, a solid thin-framed structure, known as a strongback, is glued to the window layers.  

\begin{figure}[htb]
	\begin{center}

 \begin{minipage}{6in}
  \centering
  \raisebox{-0.5\height}{\includegraphics[height=1.75in]{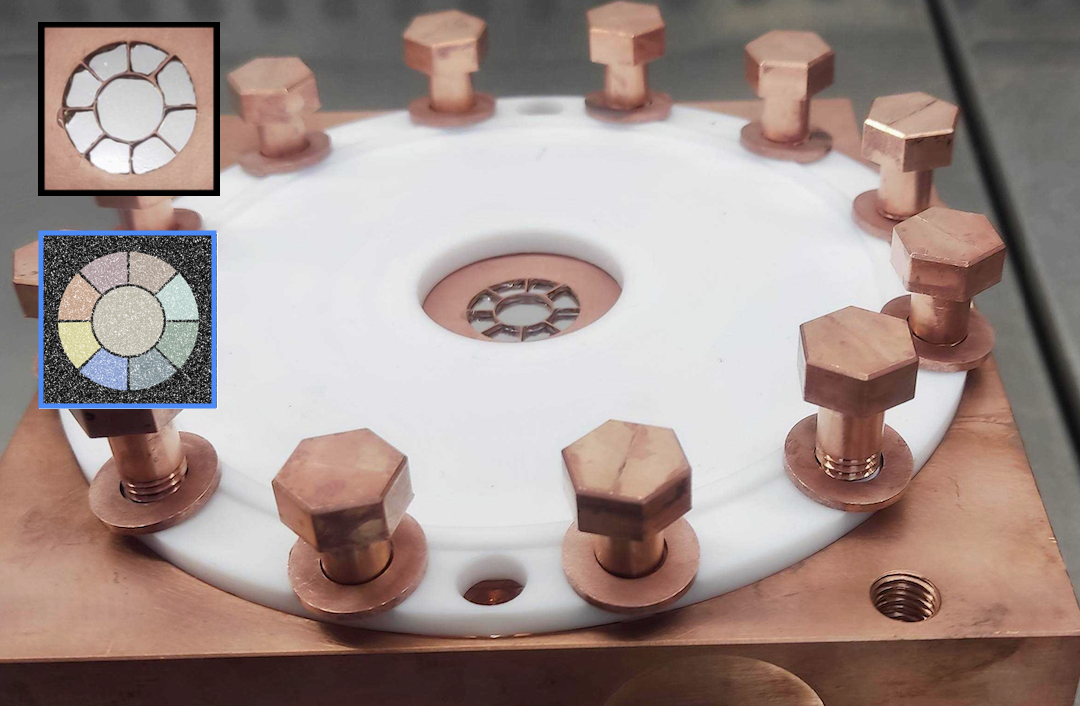}}
  \hspace*{.2in}
  \raisebox{-0.5\height}{\includegraphics[height=1.75in]{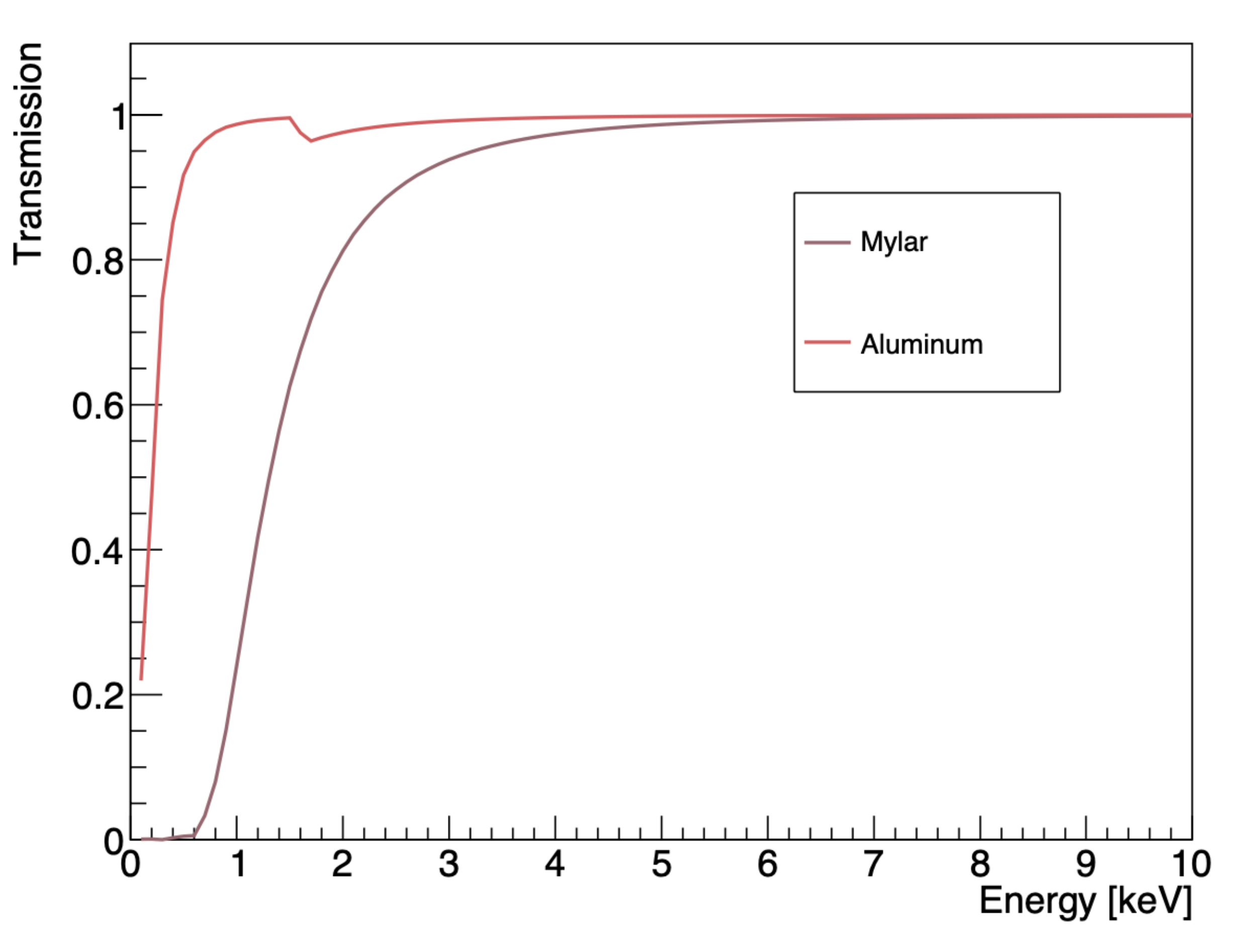}}
\end{minipage}

		\caption{On the \textit{left}, a picture of the detector-vacuum mechanical interface containing the differential window. The top-left corner shows a zoomed detail of the window, where the strongback structure and the aluminized Mylar foil are visible, along with a MC-generated image of the mask that reproduces the window strongback. The mask was generated using the \emph{PatternMask} class. Each region is drawn in different colors to illustrate the mask's capability to identify the different regions through which the photons passed. On the \textit{right}, the effective photon transmission of the window layers, 40\,nm aluminum foil and \SI{3.5}{\micro\m} Mylar foil.   }

		\label{fig:window}
	\end{center}
\end{figure}

A differential window for the Micromegas detection line has already been designed and constructed (see Figure~\ref{fig:window}). A mechanical support holds the window and allows it to be inserted into the vacuum pipeline. This support, including the strongback, is made from radiopure copper. The strongback structure frame, \SI{200}{\micro\m} thick, is designed to minimize opacity while enhancing the window's robustness, which is enclosed in a circle with an 8.5\,mm radius. The foil glued to the strongback consists of a 40\,nm aluminum layer on top of a \SI{3.5}{\micro\m} Mylar layer. 

To construct an X-ray window within the software implementation, the different layers must be defined separately using the \emph{XrayWindow} class. The transmission of the different layers will be combined during the ray-tracing stage. Each layer may define a patterned mask; in this case, only the transmission of photons that hit the mask pattern will be calculated. Photons that do not hit the pattern will have a transmission equal to 1 (e.g.\ the copper strongback). If no mask pattern is defined, the full window area will effectively contribute to the optical transmission (e.g.\ Mylar and aluminum foils). In both cases, transmission is calculated based on the thickness and material used to define the window layer. Photon absorption from solid materials used in the window construction is derived from the Henke database\,\cite{Henke:1993eda} for the relevant energy range.

\subsubsection{Micromegas window efficiency estimation}

The Micromegas window design will be used as a reference for estimating the sensitivity later in \cref{sc:sensitivity}. Here, we estimate the average window transmission, first considering the transparency of the strongback structure. Assuming that all photons are stopped by the copper strongback structure, its geometrical shape leads to the following expression:

\begin{equation}
    T_\text{sb} \simeq \frac{1}{R_\text{out}^2}\left[(1-\frac{n_{a}w_{a}}{2\pi})(R_\text{out}^2-R_\text{in}^2)+{R^2_o}\right]
\end{equation}

\noindent where $R_\text{out}=8.5$\,mm is the radius of the window boundary, $ R_\text{in}=4.55$\,mm is the outer radius of the inner circle, $R_o$ is the inner radius of the inner circle, while $n_a=8$ and $w_a=2.64^\circ$ are the number of strongback arms and their corresponding angular width\,\footnote{These parameters can also be found on the windows file, \emph{windows.rml}, inside the axionlib data repository, library release 2.4.}.

The average contribution of the aluminum and Mylar foils to the optical transparency is convolved with the solar axion Primakoff flux as a function of the energy, similar to the method used in \cref{eq:Reff} for optical reflectivity. The contributions from each component---strongback, aluminum and Mylar---are combined to provide the final efficiency of the Micromegas X-ray window:

\begin{equation}
    \epsilon_w = T_\text{sb} \times T_\text{aluminum} \times T_\text{Mylar} =  0.92 \times 0.99 \times 0.89 = 0.82.
\end{equation}

%% file: sections/03_sensitivity.tex
\section{From ray-tracing to helioscope sensitivity}
\label{sc:sensitivity_calculation}

\begin{figure}[ht]
	\begin{center}
		\includegraphics[width=0.98\linewidth]{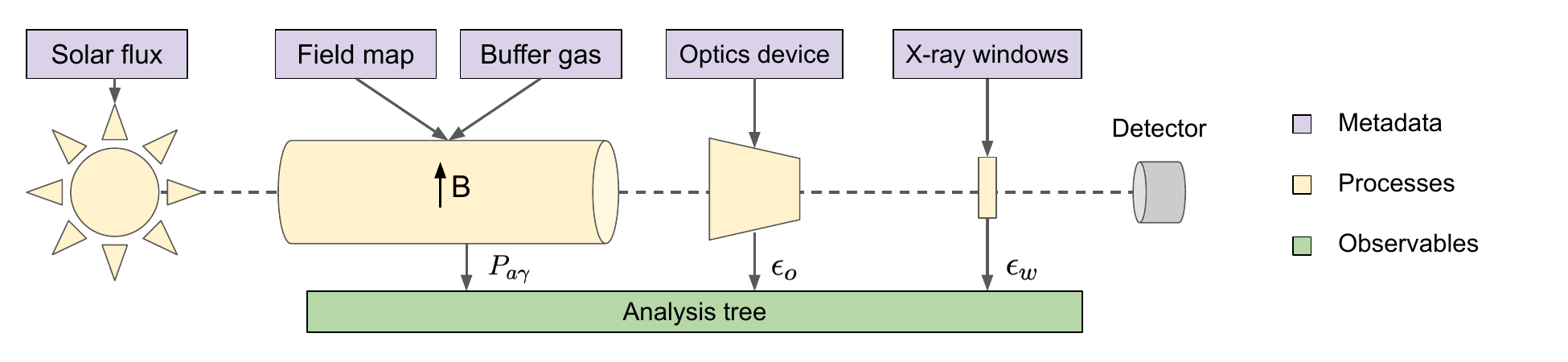}
		\caption{A schematic of the four main processes in the ray-tracing event processing chain: \emph{Generator}, \emph{FieldPropagation}, \emph{Optics}, and \emph{Transmission}. Metadata components correspond to those described in \cref{sc:components}. All processes, except for \emph{Generator}, add observables to the analysis tree, such as the axion-photon conversion probability, $P_{a\gamma}$, the optics efficiency, $\epsilon_o$, and the windows transmission efficiency, $\epsilon_w$, for each event.}
		\label{fig:rayTracing}
	\end{center}
\end{figure}

The helioscope components described in the previous section operate as standalone modules, each providing access to specific calculations. To construct a complete helioscope setup, we interconnect these modules using a \textit{REST-for-Physics} event processing chain\,\cite{Galan_2020}, which processes a dedicated axion event type unique to the axion library. This axion event type includes essential physical properties for each step of the ray-tracing calculation, such as position, momentum, axion mass, and energy. The processing chain involves several dedicated processes: \emph{Generator}\footnote{In the text, process names such as \emph{Generator} translate to class names like \emph{TRestAxionGeneratorProcess} in the C++ code.}, \emph{FieldPropagation}, \emph{Optics} and \emph{Transmission}. Calculations performed within these processes are recorded in the ROOT analysis tree, which contains all relevant information for subsequent analysis (see Figure\,\ref{fig:rayTracing}). Each process takes the position and momentum of the particle as input and may alter its trajectory, as illustrated by the \emph{Optics} component calculations shown in Figure~\ref{fig:optics}. Consequently, the order of the calculations in different ray-tracing steps may not commute within the processing chain.

It is important to note that \textit{axionlib} encapsulates the standard \textit{REST-for-Physics} event process, adding capabilities for applying rotations and translations to the helioscope components once integrated within a process. This functionality allows components to be positioned in physical space with arbitrary rotations. Therefore, components are designed to be perfectly centered and aligned with the helioscope axis, while the event process handles any requested misalignments. This way, we can translate and rotate any existing or future helioscope component/process using a common interface. This capability is crucial for studies that consider component misalignment in signal acceptance studies\,\cite{vonOy2024}. Nevertheless, in the present work, we will focus on studying the signal assuming perfect alignment of the components.

\begin{figure}[ht]
	\begin{center}
		\includegraphics[width=0.98\linewidth]{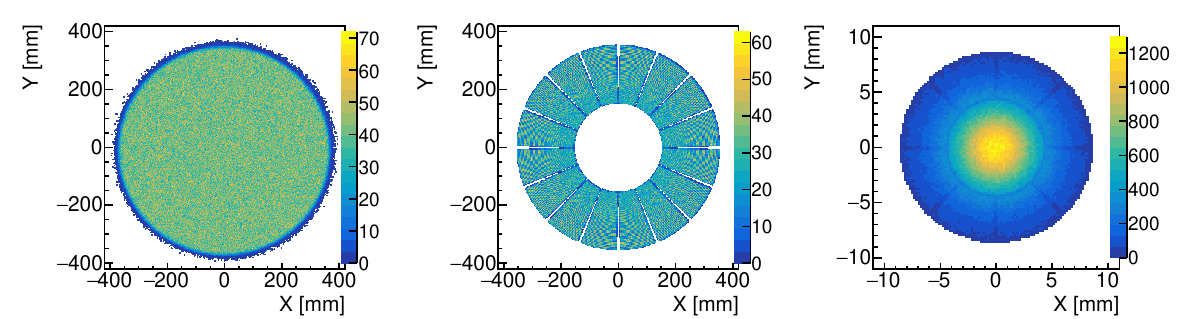}
		\caption{Spatial particle track distributions extracted at different stages of the event ray-tracing processing chain. On the left, a homogeneous particle position distribution at the end of the magnetic field map, $z=0$\,m. In the middle, the photon distribution at the X-ray optics entrance, $z=7$\,m, showing shadows from the optics spider structure. On the right, the photon distribution at the X-ray detector window, where we identify the shadows produced by the strongback structure and concentration towards the center due to focusing optics.}

		\label{fig:hitmaps}
	\end{center}
\end{figure}

The axion helioscope signal is calculated through the ray-tracing of the particles across different helioscope component processes. The \emph{Generator} process uses Monte Carlo methods to generate a particle positioned at the solar disc with a distribution defined by the solar axion model. The particle is launched toward an extensive target, which in this case is a circle matching the magnet bore aperture radius, located at the end of the magnetic field map. The \emph{FieldPropagation} process positions the center of the field map at $z=-10$\,m, meaning that the field extends from $z=-20$\,m to $z=0$\,m. This process integrates the field and calculates the axion-photon conversion probability, $P_{a\gamma}$, as described in \cref{sec:magnet}. The \emph{Optics} process, located at $z=7$\,m performs ray-tracing and applies reflectivity based on the incidence angle and photon energy, providing the optical efficiency, $\epsilon_o$, for each particle. Finally, the \emph{Transmission} process, positioned at the X-ray detector entrance (at the focal plane, at $z=\SI{7538}{\mm}$, measured from the optics location), calculates the window efficiency, $\epsilon_w$, for each particle. An additional \emph{Analysis} process may be used at various points in the chain to extract the particle's state at intermediate stages (see Figure~\ref{fig:hitmaps}).

\input{sections/03_sensitivity_montecarlo}

\input{sections/03_sensitivity_prospects}

\input{sections/03_sensitivity_statistics}

%% file: sections/03_sensitivity_montecarlo.tex
\subsection{Monte Carlo signal production}\label{sc:sensitivity}

BabyIAXO is sensitive to a wide range of axion masses. When the helioscope's magnetic field is in vacuum, the experiment can measure the axion-photon coupling, $g_{a\gamma}$, for masses up to a few tens of meV, where the axion-photon conversion in the vacuum starts to lose coherence. For higher masses, the magnetic field region might be filled with a light-weight gas, such as helium, recovering the coherence for higher masses, and extending the search towards a wider mass range.

\begin{figure}[htb]
	\begin{center}
		\includegraphics[width=0.65\linewidth]{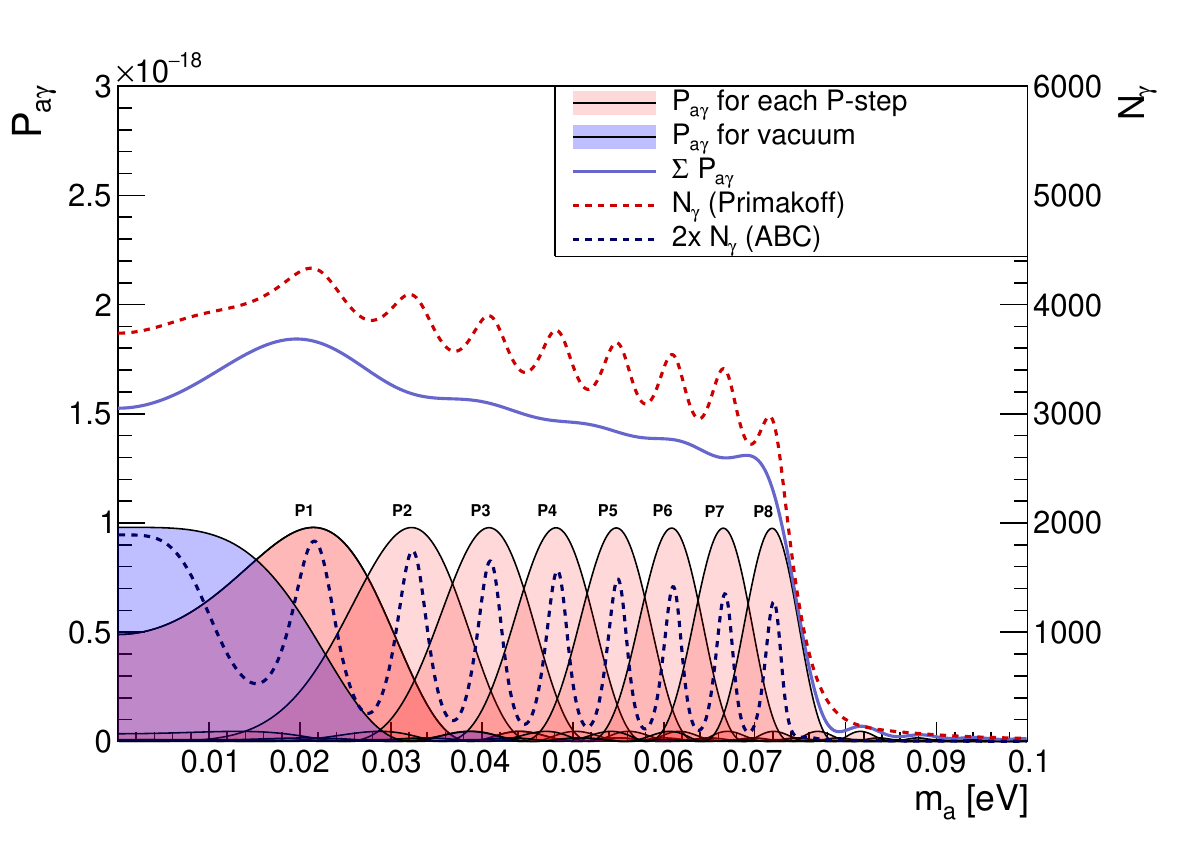}
		\caption{Axion-photon conversion probability for $g_{a\gamma}=10^{-10}$\,GeV$^{-1}$ in a constant magnetic field of 2\,T, and length of 10\,m, for a photon of energy 4.2\,keV. The probability is shown for the vacuum phase (shaded blue) and the first \emph{eight} density settings (shaded red) leading to a continuous mass scan. The integrated probability at 4.2\,keV is also shown, together with the effective number of photons expected after convolution with the axion-photon flux (Primakoff and ABC flux components). We have integrated the flux over a magnet area of  $\pi\cdot 35^2$\,cm$^2$ and an exposure time of 45$\times$12\,hours for each setting. The ABC flux intensity was convoluted assuming $g_{ae}=10^{-12}$\,GeV$^{-1}$.}
		\label{fig:resonances}
	\end{center}
\end{figure}

Introducing a buffer gas medium enhances sensitivity at a narrow mass resonance determined by the gas density, as detailed in \cref{sec:magnet}. To achieve a continuous mass scanning, these narrow resonances must overlap by gradually increasing the gas density. Figure~\ref{fig:resonances} illustrates the mass scanning strategy using discrete density settings labeled as $P_i$.

The density interval for each setting is chosen so that the next step is positioned at the full-width-half-maximum (FWHM) from the previous mass resonance when evaluated at 4.2\,keV. This energy corresponds to the peak solar axion flux rate of the Primakoff component\footnote{See also class method TRestAxionField::GetMassDensityScanning.}. The optimization for this solar flux component is evident in \cref{fig:resonances}, when comparing the expected number of photons, $N_\gamma$, from Primakoff and ABC fluxes. The modulation amplitude as a function of the mass for the expected photons produced by the Primakoff flux is lower than that expected from ABC flux, highlighting the importance of carefully tuning the scanning node densities for smoother mass scanning. This tuning ultimately depends on the energy shape of the axion signal component being measured.

Determining the sensitivity curve of BabyIAXO (see Section~\ref{sec:sensitivity}) requires calculating the axion signal through ray-tracing at each mass and density setting via MC simulations. The magnitude of the calculated signal, as described in \cref{eq:probability}, depends on both the axion mass and the gas density, which cannot be easily factorized. Consequently, each ray-tracing event must be integrated separately. The ray-tracing produces a signal distribution as a function of the detector position and energy, constructed by summing the contributions to the expected rate of each independent MC event for a given mass and density. The data workflow required to produce these signal component distributions is detailed in Appendix~\ref{sec:appendixB}.

\begin{figure}[htb]
	\begin{center}
		\includegraphics[width=0.6\linewidth]{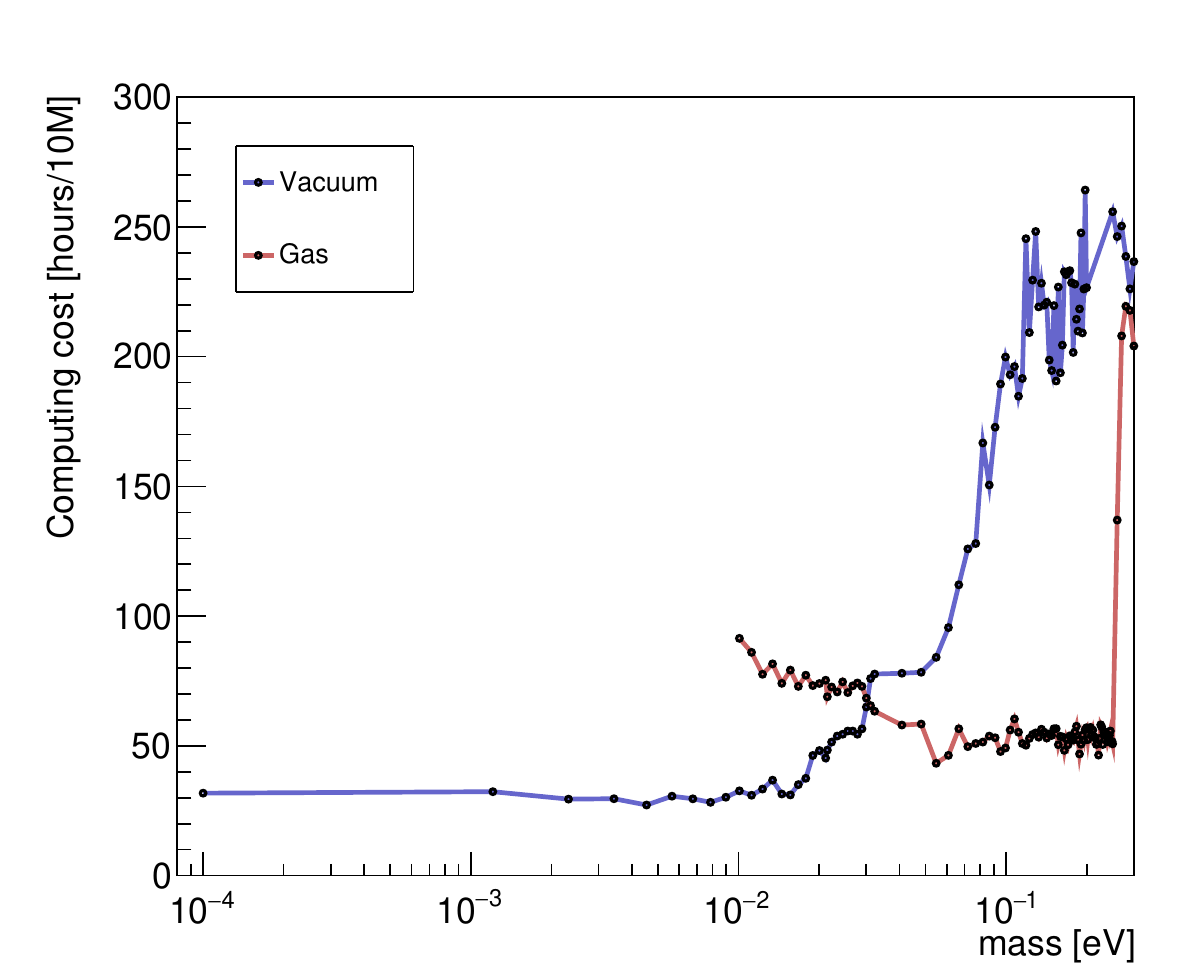}
		\caption{The required computing time normalized to 10~million events for each calculated axion mass. We differentiate the computational times required for the vacuum and gas data-taking phases. The data points corresponding to the flat minimum in both scenarios represent the case where the calculated data point is near the mass resonance. The average computing time per event starts to increase as soon as we move far away from the resonance conditions.}
		\label{fig:computingCost}
	\end{center}
\end{figure}

To ensure accurate sensitivity curves, sufficient statistics must be generated for each density setting (including vacuum) that contributes to a given mass. We will discuss in \cref{sec:statistics} that at least ten million MC events per mass and data-taking condition are necessary to reduce signal uncertainties affecting the calculated axion-photon coupling. The computation time required to produce sufficient statistics is considerable, largely driven by the computational cost of the field integration, as described in  \cref{sec:magnet}. Building the axion signal likelihood in vacuum for a total of 83 masses required approximately 25,000 CPU~hours, while the density scanning of 73 density settings, covering masses up to 0.25\,eV, required about 69,000 CPU~hours. This computation was carried out thanks to the National Analysis Facility (NAF) hosted at DESY~\cite{NAF2014}. To facilitate parallelization, these tasks were divided into approximately 2,400 and 20,800 runs, respectively, producing a total of more than 12 billion events. The MC ray-tracing presented in this work was performed with full field map accuracy of 10$\times$10$\times$50\,mm$^3$ and a tolerance of 0.01 (see \cref{tab:grid,tab:tolerance}). Note that the computational cost depends on the experimental conditions under which the field integration is performed. As detailed in Section~\ref{sec:integration}, when moving far from mass resonance conditions, the numerical method for field integration must solve a highly oscillatory function, affecting both computation time and accuracy (see Figure~\ref{fig:computingCost}).



%% file: sections/03_sensitivity_prospects.tex
\subsection{Sensitivity prospects}\label{sec:sensitivity}

A tentative physics program for BabyIAXO involves dividing the data collection into two phases, each with an exposure time of 1.5~years with 12 hours of effective solar tracking per day\,\cite{IAXO:2019mpb,IAXO:2020wwp}. The first phase would take data with a vacuum inside the magnetic field volume, while in the second phase a density scanning program following the strategy described in the previous section would take place. In order to reach the mass range up to 0.25\,eV we require to setup at least 73 discrete density settings. The exposure time, $t_\text{exp}$, at each of those settings has been distributed such that $t_\text{exp}\propto m^{-4}$, which guarantees the sensitivity curve will follow the theoretical prediction of the KSVZ line. The total time required for the gas phase to reach this benchmark at each mass resonance exceeds the 1.5-year constraint. Therefore, we reduce the time allocated to the first five density settings---those requiring longer exposure times to reach KSVZ---and redistribute it equally among them. A list with the exposure times used at each density setting is given in \cref{sec:appendixD}.

The signal prediction from the ray-tracing simulation does not account for the detector response since the detector's quantum efficiency typically does not depend on the photon's position or direction. The energy response matrix can thus be convolved with the signal prediction in a later step.

\begin{figure}[hbt!]
	\begin{center}
		\includegraphics[width=0.55\linewidth]{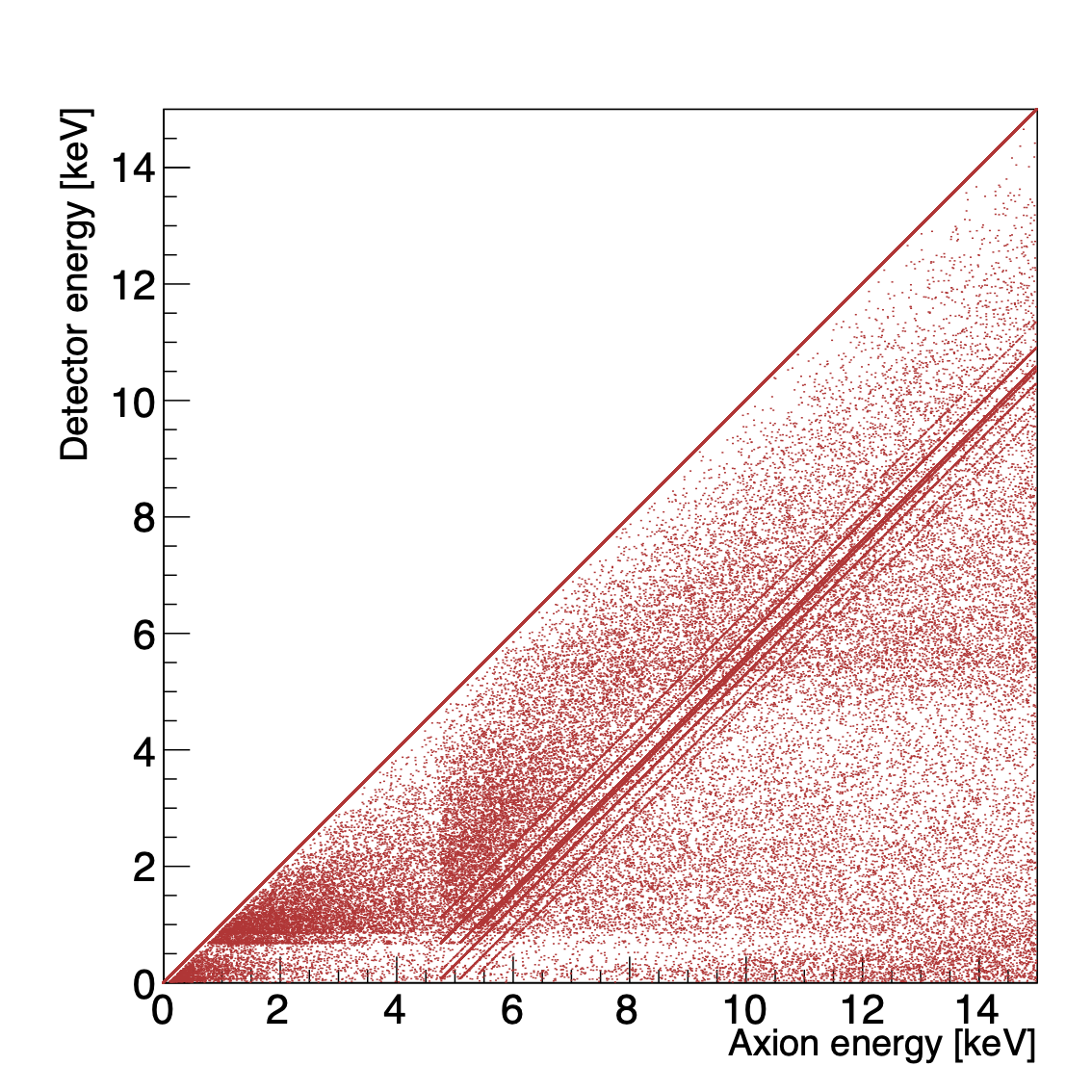}
		\caption{Photon detection response showing the detector energy measured as a function of the initial axion-photon incident energy. Most photons detected fall on the bisector line, where the incident photon is fully absorbed by the photoelectric effect.}
		\label{fig:detectorResponse}
	\end{center}
\end{figure}

In this study we considered the response matrix of a gaseous TPC with a 3\,cm drift, simulated using \emph{restG4}\footnote{See example \#14 in the restG4 repository\,\cite{luis_antonio_obis_aparicio_2024_12804201}.} for a gas mixture of 50\% xenon and 50\% neon at 1.4 bar, measured in volume (see Figure~\ref{fig:detectorResponse}). This is a potential gas mixture under investigation for the micromegas detection line\,\cite{Altenmueller:2024aak}. The resulting overall detector efficiency, $\epsilon_d$, for this xenon-neon mixture is 0.66 when integrated over the range from 0.5\,keV to 7\,keV. The detector response matrix is implemented within the \textit{REST-for-Physics} sensitivity classes. These classes provide the necessary interfaces to integrate the different experimental conditions required for the calculation - including background, signal, tracking counts, and the detector response - and combine different data taking periods and detection lines through a likelihood analysis (see Appendix~\ref{sec:appendixC}).

\begin{figure}[ht!]
	\begin{center}
		\includegraphics[width=0.7\linewidth]{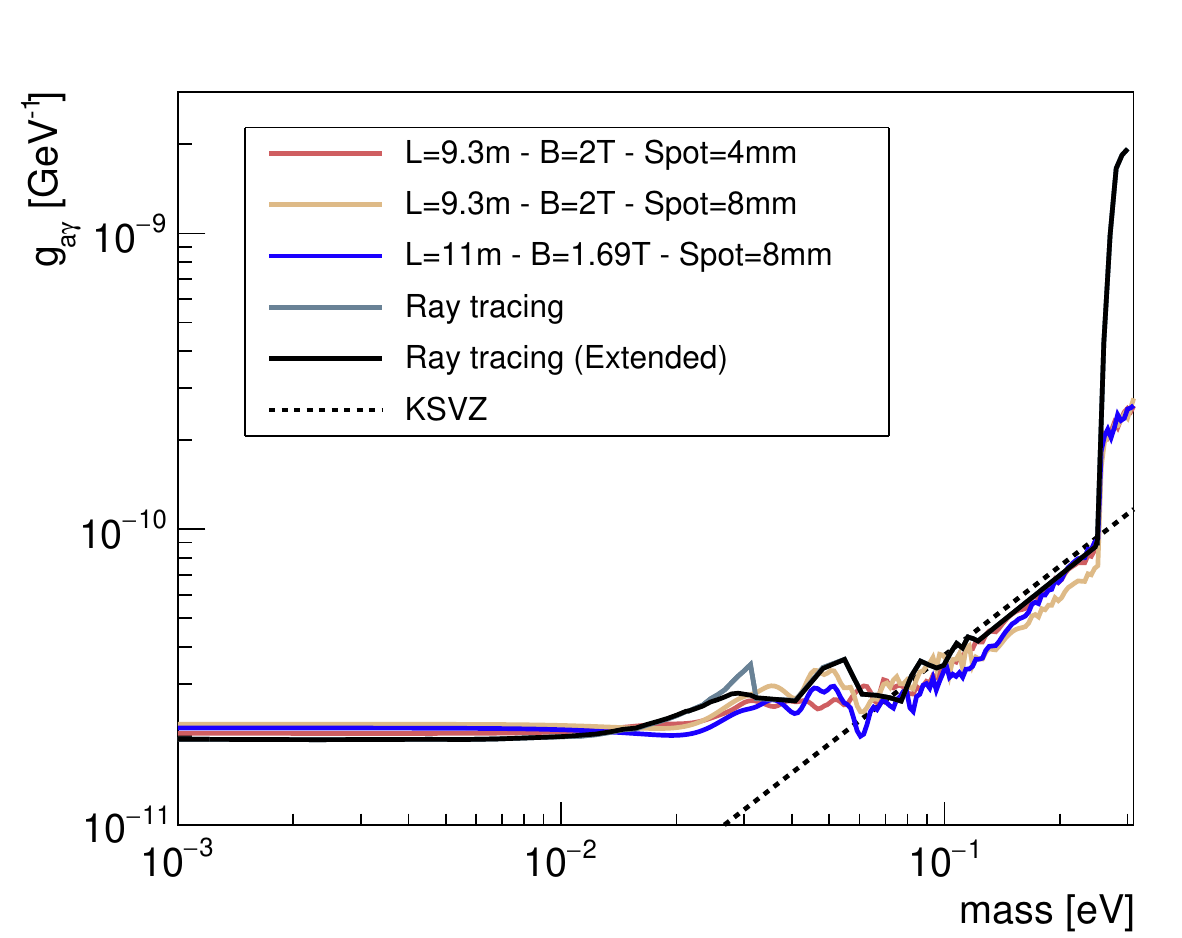}
		\caption{Sensitivity curves generated using the ray-tracing signal components for the scenarios described in the text. The "extended ray-tracing" version refers to the sensitivity curve where a larger number of mass nodes have been included in the first density settings, in the 10\,meV to 50\,meV range. For comparison, conventional calculations using a uniform field and constant coherence length under different scenarios are shown. For those cases, we use the average values obtained for the optical and windows efficiencies from Section~\ref{sc:components}, $\epsilon_o=0.47$ and $\epsilon_w=0.82$, and the overall detector response given in this section, $\epsilon_d=0.66$. }
		\label{fig:RTSensitivity}
	\end{center}
\end{figure}

\sloppy The sensitivity curves generated (see Figure~\ref{fig:RTSensitivity}) assume a flat spectrum background flux of 10$^{-7}$\,keV$^{-1}$cm$^{-2}$s$^{-1}$ and use the corresponding average for the number of simulated tracking counts at each data taking period. The signal was produced by adding the contributions from two detection lines with the same characteristics. We have defined 77 mass nodes where the experimental signal was calculated, covering masses up to 0.25\,eV. In the vacuum phase, the signal component was calculated using ray-tracing for each mass node, while in the gas phase, it was calculated only at the mass resonance of each density setting and one additional mass node on either side of the resonance.

The axions signal contribution to the sensitivity for the first density settings is wider in mass range than that for settings at higher masses. Therefore, we need to calculate the signal for a greater number of mass neighbors in the first density settings, as their contribution in a reasonable number of mass nodes is not negligible. 
An extended ray-tracing MC simulation was carried out to include the contribution of the first five density settings for masses between 10\,meV and 50\,meV, which required 22 additional mass nodes per setting.
 
The ray-tracing sensitivity achieved is comparable to that obtained using the conventional helioscope signal calculation\footnote{See example \#3 found in the axionlib repository~\cite{jgl}.}, where a homogeneous magnetic field and a uniform event distribution at the detector focal plane are assumed. In this conventional approach, the two main sources of uncertainty are the estimated magnetic field coherence length and the assumed focused spot area of the optics. As shown in Figure~\ref{fig:RTSensitivity}, different magnetic lengths, satisfying $B^2L^2\sim340$\,T$^2$m$^2$, and different fiducial spot sizes have been considered, including a spot size of 4\,mm with 50\% efficiency and a spot size of 8\,mm with 90\% efficiency. The resulting sensitivity curves provide an idea of the uncertainties in the calculation. Additionally, it should be noted that the ray-tracing signal calculation fails to reproduce the sensitivity in regions outside the mass resonance, as observed for axions masses above 0.25\,eV, due to the numerical integration errors discussed in Section\,\ref{sec:integration}.

%% file: sections/03_sensitivity_statistics.tex
\subsection{Impact from background and signal uncertainties}\label{sec:statistics}

\begin{figure}[b]
	\begin{center}
         \includegraphics[width=0.66\linewidth]{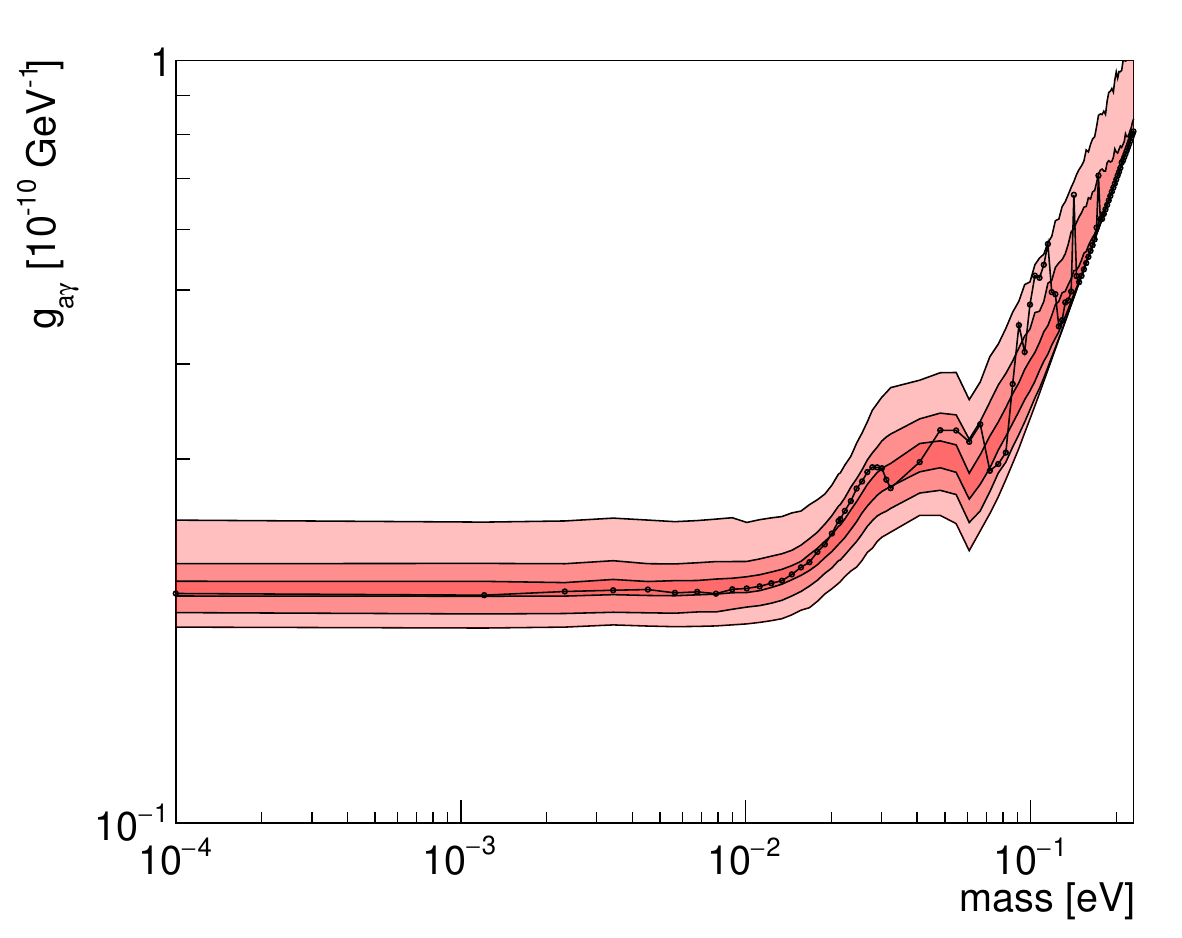}
		\caption{Contour levels showing the likelihood of BabyIAXO achieving a given sensitivity in $g_{a\gamma}$. The darkest red area represents the area where 25\% of the experiments fall at each given mass, while the lighter red regions correspond to the 68\% and 95\% confidence levels, respectively. The black data points represent the sensitivity curve generated by a randomly selected single-experiment MC simulation.}
		\label{fig:Bsystematics}
	\end{center}
\end{figure}

The sensitivity curve for BabyIAXO in Figure~\ref{fig:RTSensitivity} used the expected number of counts and ignored statistical fluctuations of the data. Here we remedy this by performing MC simulations of mock datasets.  Figure~\ref{fig:Bsystematics} shows the confidence levels for the BabyIAXO sensitivity prospects obtained from about 1,000 simulated experiments, covering both vacuum and gas phases. In the vacuum phase, the central value for $g_{a\gamma}$ is
\[
    g_{a\gamma}=2.03^{+0.16}_{-0.14}\cdot10^{-11}\,\text{GeV}^{-1} \quad \text{(68\% C.L.).}
\]

As seen in Figure~\ref{fig:Bsystematics}, the fluctuations associated with a single experiment during the gas phase are significantly larger than those obtained using the average number of tracking counts per density setting (see Figure~\ref{fig:RTSensitivity}). This is due to the arbitrary number of tracking counts measured across different density settings. Nevertheless, during the data-taking phase of the experiment, these fluctuations might be minimized by increasing the exposure time for settings where a higher number of tracking counts are observed. If the higher-than-expected count rate does not decrease to the expected level with increased exposure, the significance of the observed signal could potentially lead to a 5-sigma discovery.

\begin{figure}[hb]
	\begin{center}
		\includegraphics[width=0.66\linewidth]{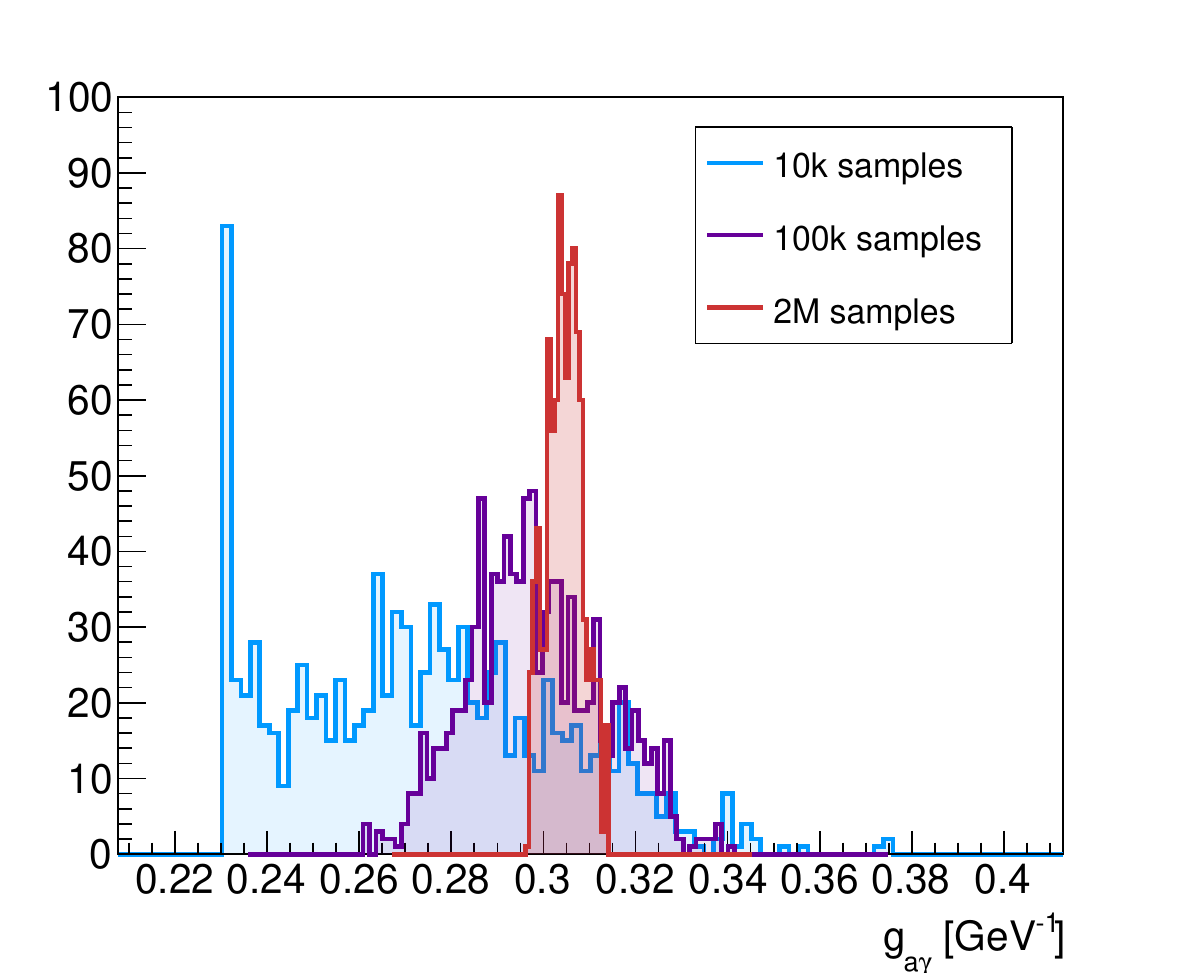}
		\caption{Distribution of $g_{a\gamma}$ calculated under vacuum conditions for a 50-day tracking period during which 15 tracking counts were observed. The signal position and energy distributions are reconstructed for each calculation using a random sample from the ray-tracing simulation. Distributions for the 10\,k, 100\,k, and 2\,M cases are shown. }
		\label{fig:Ssystematics}
	\end{center}
\end{figure}

The reconstruction of the signal component is also a source of uncertainty on the calculation of $g_{a\gamma}$. To study its impact on the axion-photon coupling, we reconstructed the axion signal distribution using different statistical sample sizes. For this purpose, we prepared a large MC ray-tracing sample of approximately 10$^9$ events, simulating a single mass node under vacuum data-taking conditions. We then randomly extracted sub-samples of various sizes -- 10\,k, 50\,k, 100\,k, 500\,k and 2\,M --  and used these to reconstruct the signal. For each sample size, we generated a distribution of the calculated $g_{a\gamma}$ (see Figure~\ref{fig:Ssystematics}).

 We tested different tracking count configurations compatible with a background level of 10$^{-7}$\,cm$^{-2}$\,s$^{-1}$\,keV$^{-1}$ using the signal reconstruction technique based on various sub-samples. The results are presented in Table~\ref{tab:signal_statistics}. Configurations \textbf{A} and \textbf{B} consist of two different position and energy distributions of 15 tracking counts over a 75-day tracking period, while configuration \textbf{C} was generated using a 10-day tracking period with only 2 tracking counts observed. Each tracking day corresponds to an effective exposure of 12 hours. As previously noted in Figure~\ref{fig:Ssystematics}, the uncertainty related to statistical signal reconstruction is significantly reduced as the sub-sample size increases, with all three configurations displaying similar behavior. It is important to highlight that all the ray-tracing sensitivity prospects presented in this work were obtained using at least 20 million events per MC simulation setting.

\begin{table}[!ht]
\centering
\begin{tabular}{>{\centering\arraybackslash}p{3cm}cccccc}
\toprule
\multirow{2}{=}{\centering\arraybackslash \textbf{Number of sub-samples}} & \multicolumn{2}{c}{\textbf{A}} & \multicolumn{2}{c}{\textbf{B}} & \multicolumn{2}{c}{\textbf{C}} \\
\cmidrule{2-3} \cmidrule{4-5} \cmidrule{6-7}
& \textbf{Mean} & \textbf{Sigma} & \textbf{Mean} & \textbf{Sigma} & \textbf{Mean} & \textbf{Sigma} \\
\midrule
10,000     & 2.74  & 0.29 & 2.46 & 0.21 & 3.88 & 0.10 \\
50,000     & 2.93  & 0.20 & 2.57 & 0.19 & 3.89 & 0.12 \\
100,000    & 2.98  & 0.15 & 2.60 & 0.17 & 3.90 & 0.12 \\
500,000    & 3.04  & 0.08 & 2.65 & 0.09 & 3.92 & 0.09 \\
2,000,000  & 3.05  & 0.04 & 2.66 & 0.05 & 3.92 & 0.06 \\
\bottomrule
\end{tabular}
\caption{Statistical mean and standard deviation (sigma) of the axion-photon coupling distribution, in units of 10$^{-11}$GeV$^{-1}$. Each value is derived from 1,000 different experiments conducted under the same experimental conditions (vacuum phase), updating the signal reconstruction sub-sample for each calculation. Results for different experimental conditions, \textbf{A}, \textbf{B}, and \textbf{C}, as described in the text, are shown.}

\label{tab:signal_statistics}
\end{table}

%% file: sections/05_conclusions.tex
\section{Conclusions}
\label{sec:conclusions}
We reviewed the fundamentals of calculating the axion signal in BabyIAXO to provide an accurate estimate of the signal likelihood distribution in both position and energy, as observed by the experiment's X-ray detectors.
Precisely computing the signal is essential for determining the experimental sensitivity and prospects. To achieve this, we have developed \textit{axionlib}, a dedicated library in \textit{REST-for-Physics} that incorporates the various helioscope components affecting axion-photon propagation inside the apparatus. The motivation behind this work is to establish a benchmark analysis pipeline for IAXO.

For the first time, we have tackled the challenge of calculating the axion-photon conversion probability within an inhomogeneous magnetic field in axion helioscopes. This is a significant advancement that enables the calculation of the axion signal using Monte Carlo simulations through the ray-tracing technique. Computing the axion-photon conversion in an inhomogeneous magnetic field is computationally intensive and represents a bottleneck in signal computation---a challenge that we have thoroughly addressed in this work, including error estimations under various conditions.

We included an idealized optical response that accounts for the detailed mirror geometry and the properties of the reflecting layers. However, additional effects, such as diffuse reflection, mirrors deformation, or as-built metrology were not included but could be implemented in future studies. Moreover, the final sensitivity of the experiment could benefit from incorporating an experimentally measured optical response, encoded in a comprehensive optics response matrix. This matrix could then be directly integrated into a \textit{REST-for-Physics} process to produce a more realistic representation of the apparatus's response.

We have demonstrated that the sensitivity levels obtained from ray-tracing signal calculations are comparable to those from simplified homogeneous magnetic field models. However, this does not imply that the simplified calculation is sufficiently accurate. Variations in sensitivity curves arise even within different scenarios of the simplified calculation, such as changes in the magnetic length or the optics focal spot size, resulting into uncertainties based on our assumptions about these parameters. Obtaining an accurate sensitivity curve thus requires ray-tracing simulations that account for all features of the magnetic field map, in combination with the other helioscope components.

We focused on calculating the sensitivity prospects of BabyIAXO solely for the Primakoff solar axion component. A more comprehensive analysis should include additional components, such as the ABC axion-electron flux, the \ce{^57Fe} axion-nucleon component, or axion-photon production mechanisms in macroscopic solar magnetic fields. The \textit{REST-for-Physics} library provides a complete toolkit for studying these various solar components and facilitates comparative studies in a unified framework. 
Thanks to being publicly available, its outputs---such as the BabyIAXO instrument response function---is accessible to the community for studying the potential of solar axion searches.

In summary, this work represents a significant milestone for BabyIAXO by providing a reference for future calculations. It facilitates the evaluation of various helioscope component modifications, such as new optics or magnetic field configurations.
Finally, it establishes the \textit{REST-for-Physics} axion library we developed as a promising platform for testing and refining axion helioscope searches.

%% file: sections/appendixA.tex
\section{Optics interface}\label{sec:appendixA}
\begin{figure}[ht]
	\begin{center}
		\includegraphics[width=0.6\linewidth]{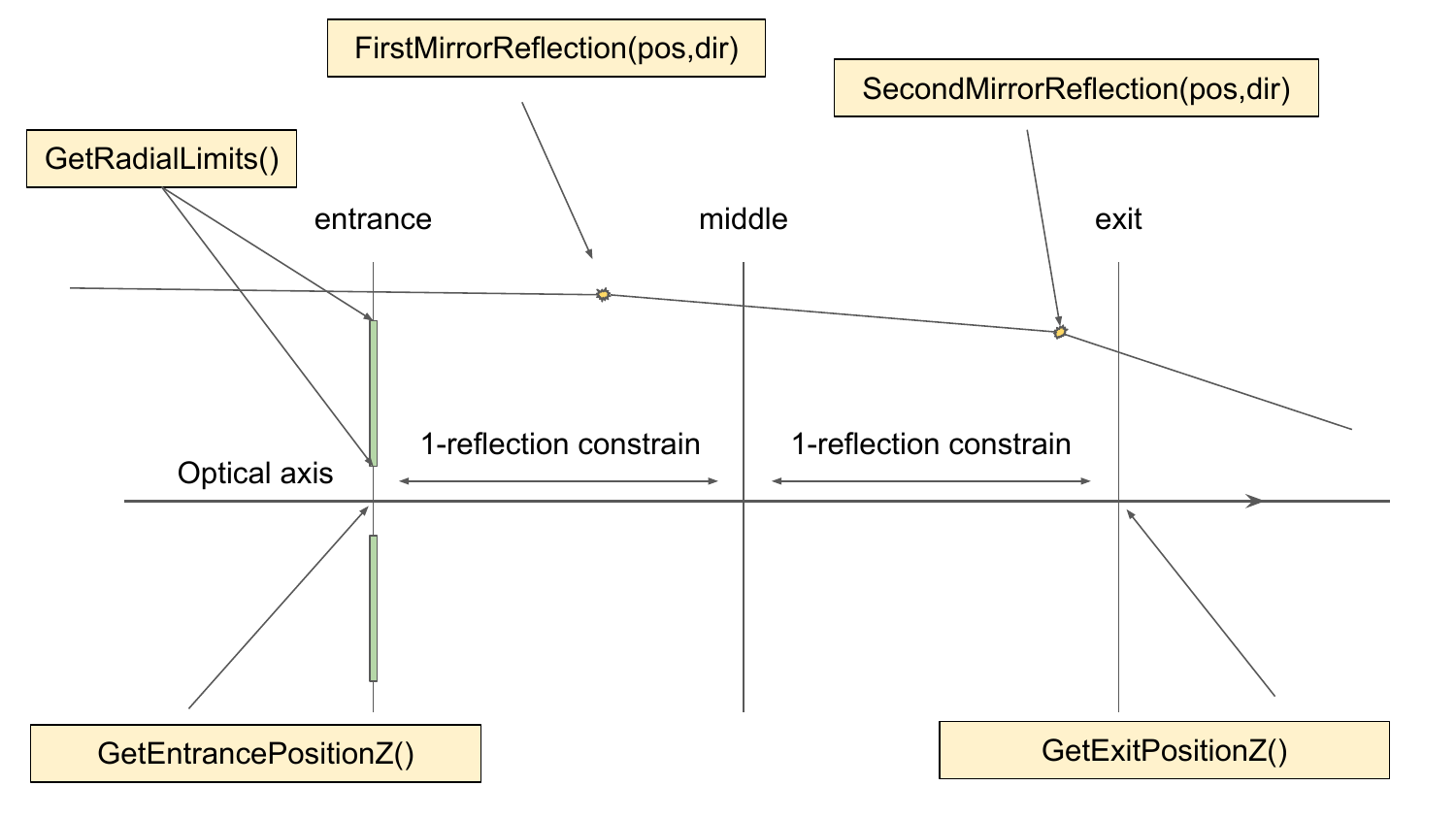}
		\caption{Schematic of the software implementation of geometrical optics within the \emph{Optics} class. Yellow boxes represent pure virtual methods that must be implemented by any derived optics class.}
		\label{fig:opticsInterface}
	\end{center}
\end{figure}

The \textit{REST-for-Physics} \emph{Optics} event process implemented within \textit{axionlib} uses the \emph{Optics} metadata class to describe focusing devices. This class is a pure abstract class that defines a common pattern or interface for implementing various specific optical geometries. It defines common methods that must be implemented by any specific optics implementation inheriting from this class. Additionally, it defines three virtual planes (entrance, middle, and exit) that serve to control and constrain the movement of particles inside the device (see Figure~\ref{fig:opticsInterface}). The derived class is responsible for providing all the necessary details to describe a particular optics device, such as the \emph{TrueWolterOptics} class used in this work. These details include the mirror's geometrical parameters and properties, as well as the physics of the reflection specific to that particular focusing device, like the particle reflection unique to a given shape of the mirror\footnote{Basic geometrical reflections, such as conic, hyperbolic or parabolic reflections, can be found in the \emph{REST\_Physics} namespace along with other basic vector operations.}.

\begin{figure}[ht]
	\begin{center}
		\includegraphics[width=0.48\linewidth]{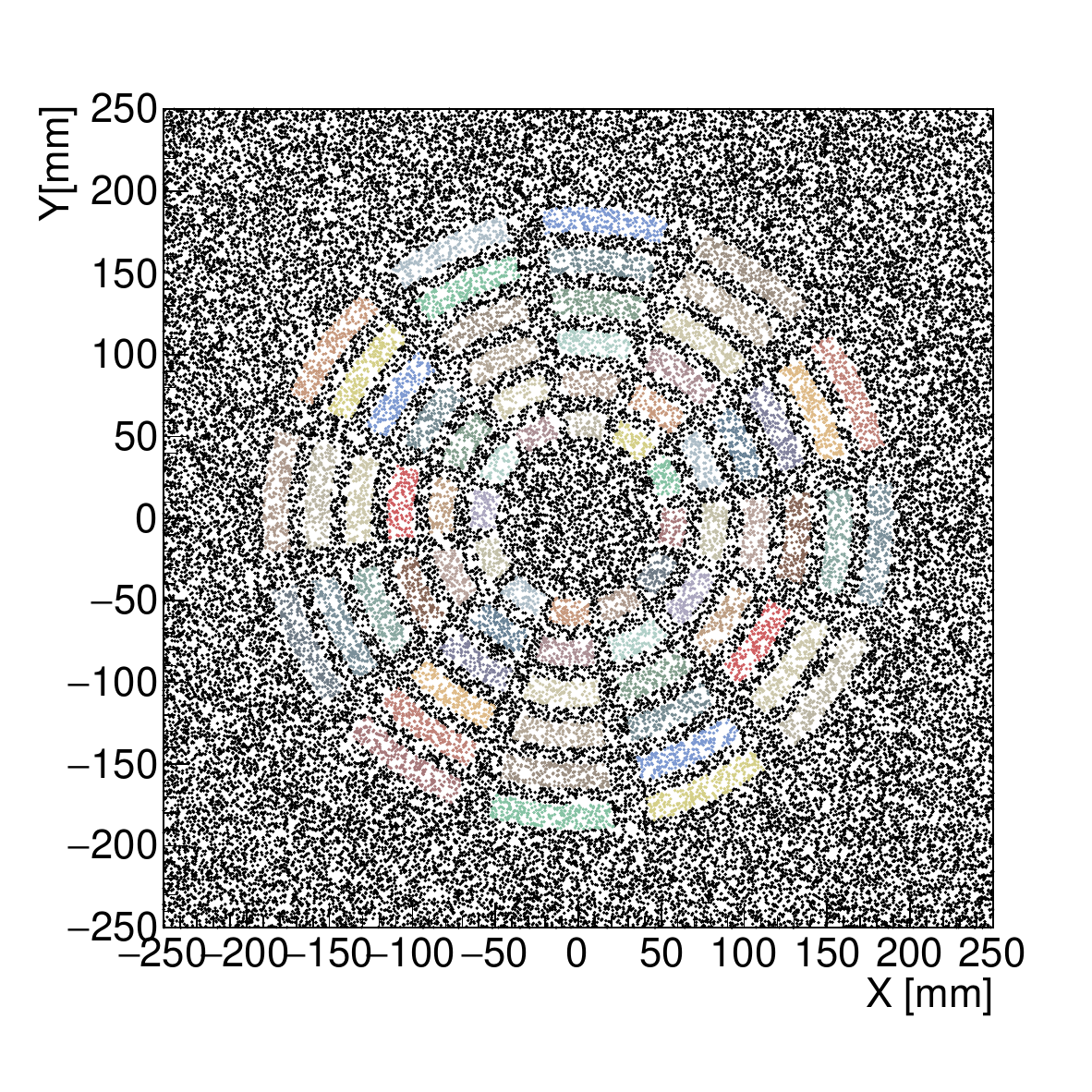}
		\caption{A dummy mask pattern illustrating the result of a \emph{RingMask} and \emph{SpiderMask} class built through the \emph{CombinedMask} class. This illustration was produced using a Monte Carlo simulation. Different colors represent regions with unique IDs, while black dots represent areas where no region is found. }

		\label{fig:ringMasks}
	\end{center}
\end{figure}

The three virtual planes serve to constrain the direction of photon propagation from the entrance to the exit plane. For a photon to be successfully focused, it must be found ahead of the entrance plane and moving towards it. The \textit{REST-for-Physics} framework includes a set of classes capable of constructing two-dimensional masks with various patterns, allowing us to identify specific regions (see Figure~\ref{fig:ringMasks}). Each of the three virtual planes has a different mask implementation. Photons passing through a specific region on the entrance mask, identified by a unique ID number, must necessarily pass through the corresponding region with the same ID on the middle and exit masks. Obviously, the masks at each inter-face should be designed to match the geometric constrains of each specific optical device.

The photon propagation is implemented by the base \emph{Optics} class through a method called \emph{PropagatePhoton}, which makes use of the methods implemented at the specific derived optics classes. While the base \emph{Optics} class reads the optics data file, the derived class is responsible for interpreting the detailed geometry of the mirrors, including the boundaries where photon reflectivity is allowed. An additional constrain ensures that the number of reflections on each mirror, between the entrance and middle planes and between the middle and exit planes, is exactly one.

%% file: sections/appendixB.tex
\section{Data management and workflow}\label{sec:appendixB}

\begin{figure}[ht]
	\begin{center}
		\includegraphics[width=0.98\linewidth]{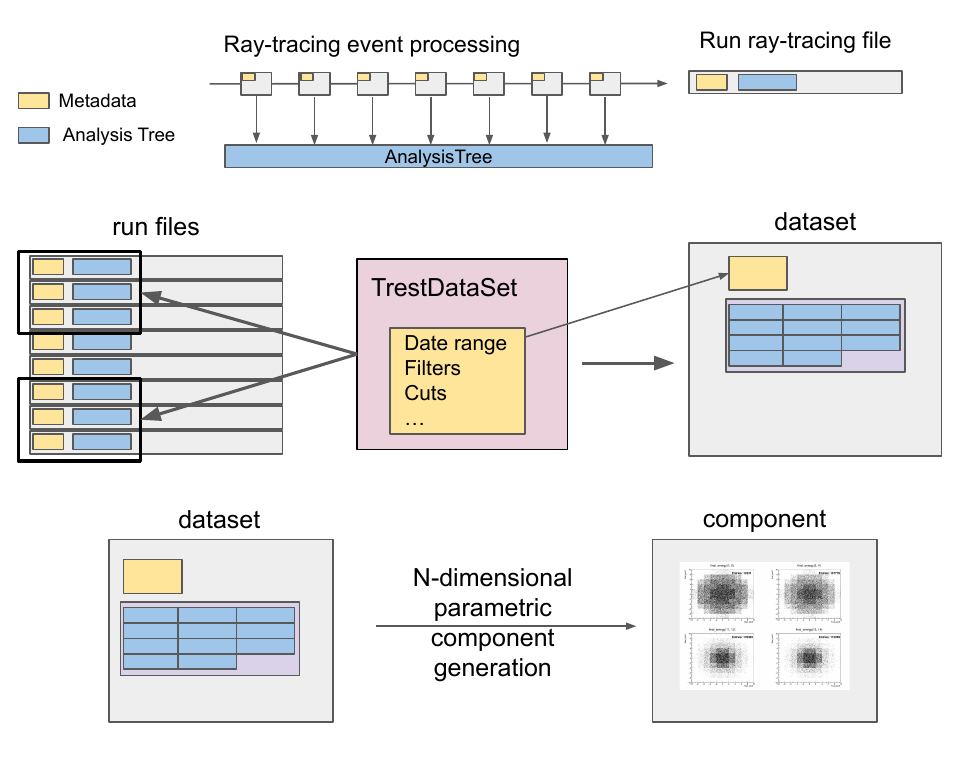}
        \caption{Data work flow for generation of an axion signal component in \textit{REST-for-Physics}.}
		\label{fig:dataWorkFlow}
	\end{center}
\end{figure}

The axion signal ray-tracing is conducted by implementing a Monte Carlo generator process within a \textit{REST-for-Physics} event processing chain. This \emph{Generator} process simulates axions originating from the solar disk following the spatial and energy distributions of a specific solar model, and directs them toward and extensive target. Photons then pass through various helioscope components, each encoded as a process that translates its weight on the axion signal into the analysis tree for each event. The last process propagates the photons to the plane where the detector is located. As shown in Figure~\ref{fig:dataWorkFlow}, the processing chain produces a ray-tracing file that contains all the metadata information used during the file's generation, as well as an analysis tree with relevant data for the final analysis. This includes the $x$ and $y$ positions at the X-ray detector plane, the final photon energy, the axion mass, and the contribution to the signal weights from each helioscope component.

The ray-tracing calculations for a specific axion mass and experimental setting are divided into parallel jobs, with each job producing an independent run file. Once all the run files have been generated, they are merged using the \emph{Dataset} class, which allows for the selection of run files based on particular metadata criteria. In this work, we compiled a single dataset containing all simulated masses for each measurement condition, whether in a vacuum or under any of the simulated density settings. The \emph{Dataset} class provides access to a unified ROOT tree and/or ROOT RDataFrame that contains all the simulated entries. Additionally, the \emph{Dataset} class can store some quantities derived from any existing metadata available inside the run file, such as the total number of simulated entries (which may differ from the number of entries in the combined tree), or the integrated solar axion flux, $\Phi_a$, and generator area, $A_g$, extracted from the \emph{Generator} process, which are crucial to determine the final event rate in the detector.

During dataset generation, new columns (or branches in ROOT terminology) can be added based on the available dataset quantities and pre-existing columns. This enables the calculation and inclusion of the event rate into the dataset, specifically the contribution of each Monte Carlo event to the overall event rate of the detector, which is calculated as follows:

\[
    n_{\gamma} = P_{a\gamma}  \Phi_a A_{g} \prod_i \epsilon_i
\]

\noindent where $P_{a\gamma}$ represents the axion-photon conversion probability calculated by the \emph{FieldPropagation} process and $\epsilon_i$ accounts for the efficiency contributions from various helioscope components, such as the X-ray window transmission or the optics reflectivity.

Once a dataset has been compiled for each measurement condition, the \emph{ComponentDataSet} class is used to generate N-dimensional density maps, referred to as components, based on the dataset's columns (see Figure~\ref{fig:dataWorkFlow}). In this work, we partitioned our signal using the final detector event position and energy, with a spatial resolution of 100\,$\mu$m, and an energy resolution of 0.5\,keV, which are suitable for the characteristics of a micromegas detector. The \emph{ComponentDataSet} class enables the parameterization of the density maps based on one of the dataset columns. This approach allows us to create a density map for each axion mass.

Each cell in this N-dimensional signal description is normalized by the number of simulated entries, $N_{sim}$, per axion mass in our Monte Carlo ray-tracing simulation. The signal component rate (measured in cm$^{-2}$\,keV$^{-1}$\,s$^{-1}$) can then be evaluated for each mass at any point within the defined range (between -10\,mm and 10\,mm for both spatial dimensions, and from 0\,keV to 20\,keV for energy) at the specified resolution.

%% file: sections/appendixD.tex
\section{Density settings exposure times}\label{sec:appendixD}

For reference, Table~\ref{table:gasTimes} provides the exposure times allocated during the second data-taking phase of BabyIAXO. The distribution of the exposure time per density setting is estimated such that the sensitivity curve will follow the tendency of the KSVZ theoretical expectation. The distribution is calculated for a total exposure time of 1.5 years, 12 hours tracking per day. The constrain of 1.5 years exposure avoids the first five density settings reaching the KSVZ line, therefore, the remaining time is equally distributed between those five first settings. The steep rise of the KSVZ line as a function of the mass creates a big difference of tracking exposure between the first settings, of the order of a month, and the last settings, requiring just few hours per tracking.

\begin{table}[htb]
\centering
\begin{tabular}{cc|cc|cc|cc|cc|cc}
P1 & 438.0 & P2 & 438.0 & P3 & 438.0 & P4 & 438.0 & P5 & 438.0 & P6 & 859.0 \\
P7 & 604.0 & P8 & 445.0 & P9 & 340.2 & P10 & 267.8 & P11 & 215.5 & P12 & 177.0 \\
P13 & 147.6 & P14 & 124.8 & P15 & 106.9 & P16 & 92.5 & P17 & 80.7 & P18 & 71.1 \\
P19 & 63.0 & P20 & 56.3 & P21 & 50.5 & P22 & 45.6 & P23 & 41.3 & P24 & 37.6 \\
P25 & 34.4 & P26 & 31.6 & P27 & 29.1 & P28 & 26.9 & P29 & 24.9 & P30 & 23.2 \\
P31 & 21.6 & P32 & 20.2 & P33 & 18.9 & P34 & 17.7 & P35 & 16.6 & P36 & 15.7 \\
P37 & 14.8 & P38 & 14.0 & P39 & 13.2 & P40 & 12.5 & P41 & 11.9 & P42 & 11.3 \\
P43 & 10.8 & P44 & 10.3 & P45 & 9.8 & P46 & 9.4 & P47 & 9.0 & P48 & 8.6 \\
P49 & 8.2 & P50 & 7.9 & P51 & 7.6 & P52 & 7.3 & P53 & 7.0 & P54 & 6.7 \\
P55 & 6.5 & P56 & 6.2 & P57 & 6.0 & P58 & 5.8 & P59 & 5.6 & P60 & 5.4 \\
P61 & 5.3 & P62 & 5.1 & P63 & 4.9 & P64 & 4.8 & P65 & 4.6 & P66 & 4.5 \\
P67 & 4.4 & P68 & 4.2 & P69 & 4.1 & P70 & 4.0 & P71 & 3.9 & P72 & 3.8 \\
P73 & 3.7 & & & & & & & & & & \\
\end{tabular}
\caption{Exposure times in hours for each gas density setting, $P_i$, used to generate the BabyIAXO sensitivity prospects.}
\label{table:gasTimes}
\end{table}

%% file: sections/appendixC.tex
\section{REST-for-Physics sensitivity classes}\label{sec:appendixC}

\begin{figure}[ht]
	\begin{center}
		\includegraphics[width=0.98\linewidth]{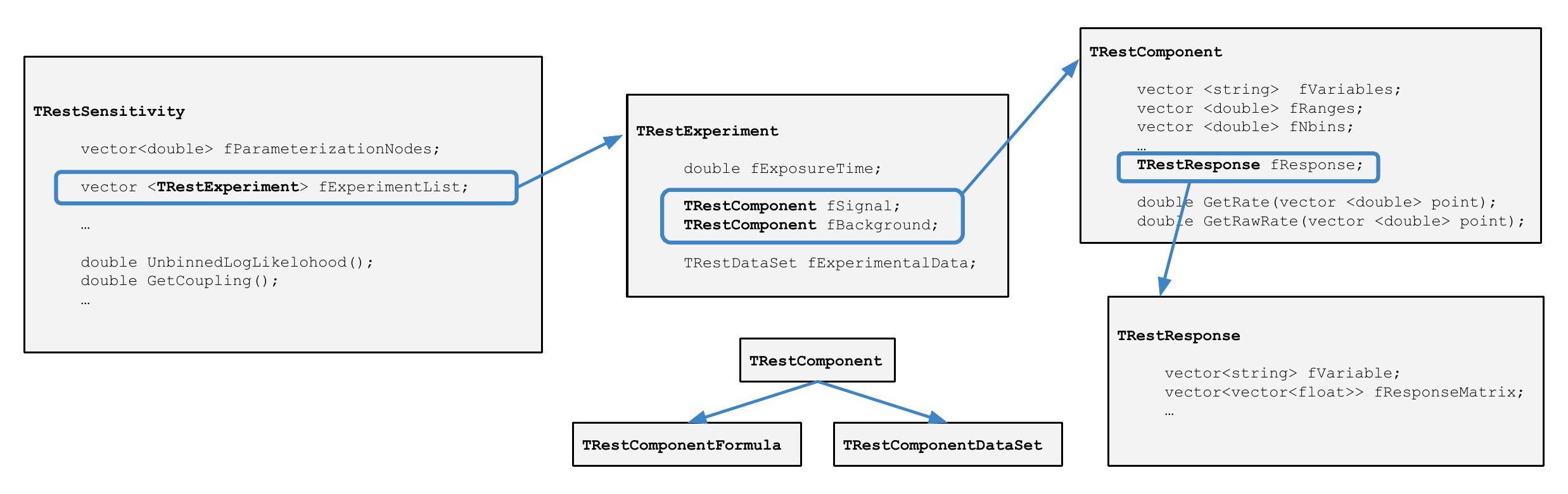}
        \caption{\textit{REST-for-Physics} sensitivity-related classes: The class descriptions have been simplified for brevity. For more detailed information, please refer to the code implementation.}
		\label{fig:sensitivityClasses}
	\end{center}
\end{figure}

The \textit{REST-for-Physics} sensitivity-related classes (see Figure~\ref{fig:sensitivityClasses}) provide interfaces for calculating the sensitivity of a data-taking period based on the detector's background, the expected axion signal, and the observed or simulated tracking data. The \emph{Component} class provides access to parametric N-dimensional density maps, facilitating the evaluation of background and signal rates within the detector's parameter space, typically $x$ and $y$ positions and energy. Initialization of such components can be done through various mechanisms. In this work, we used the \emph{ComponentDataSet} class (see Appendix~\ref{sec:appendixB}) to construct the axion signal from our Monte Carlo ray-tracing simulations and the \emph{ComponentFormula} class to model a flat background. The \emph{ComponentDataSet} can be used to describe components from either experimental or Monte Carlo datasets, enabling the incorporation of more complex background models into the sensitivity calculation, or even the use of experimental background data measured by the detectors during the data-taking phase.

The \emph{Sensitivity} class implements an unbinned likelihood method, where the likelihood $\mathcal{L}_m$ for a specific axion mass $m$ is calculated as follows\,\cite{2011arXiv1102.1406G}:

\begin{equation}
-\frac{1}{2}\chi^2(g) = \mbox{log}\,\mathcal{L}_m = g^4 \cdot \mathcal{S}^T_m \cdot t_{exp} + \sum_{Tck} \mbox{log} \,\mathcal{R}_m (x,y,E)
\end{equation}

\noindent where $\mathcal{S}_m^T$ is the total integrated axion signal rate over the full detection area and energy range, $t_{exp}$ is the measured exposure time, and the sum runs over all the observed counts under tracking conditions. The observed rate in tracking conditions, $\mathcal{R}_m$, is a direct function of the expected background rate in the detector, $\mathcal{B}$, and the axion signal contribution at a specific axion mass $\mathcal{S}_m$ as a function of position and energy:

\begin{equation}
\mathcal{R} (x,y,E) = \mathcal{B} (x,y,E) + g^4 \cdot \mathcal{S}_m (x,y,E)
\end{equation}

\noindent where the axion-photon coupling, $g$, arising from production and detection interactions has been explicitly factored out of $\mathcal{S}_m$. The resulting $g$ at 95\% C.L. is obtained when $\chi(g)=2$. Several experimental likelihoods for different experiments can be included into the sensitivity calculation to obtain a combined sensitivity:

\begin{equation}
\mathcal{L}_{total} = \prod_{Exp} \mathcal{L}_m (x,y,E).
\end{equation}

The \emph{Sensitivity} class computes this expression using a collection of experimental descriptors that leverage the \emph{Experiment} class to standardize access to background, signal components, and tracking data. The tracking data is provided inside the \emph{Experiment} class through the \emph{DataSet} class, which can originate either from experimental sources or from Monte Carlo mock-generated data. In the absence of experimental tracking data, the \emph{Experiment} class generates a Monte Carlo tracking dataset based on the background component and the predefined exposure time for the given experiment. If experimental tracking data is available, the \emph{DataSet} produced by the detector's processing chain will include the exposure time corresponding to the data in the dataset.